\documentclass[twocolumn,tighten]{aastex63}

\usepackage{graphicx}
\usepackage{ulem}
\usepackage{amsmath}
\usepackage{wrapfig}
\usepackage[thinlines]{easytable}
\usepackage{tikz}

\newcolumntype{P}[1]{>{\centering\arraybackslash}p{#1}}

\definecolor{aqua}{RGB}{4, 216, 178}

\newcommand\lsim{\mathrel{\rlap{\lower4pt\hbox{\hskip1pt$\sim$}}
        \raise1pt\hbox{$<$}}}
\newcommand\gsim{\mathrel{\rlap{\lower4pt\hbox{\hskip1pt$\sim$}}
        \raise1pt\hbox{$>$}}}

\begin{document}
\shorttitle{White dwarfs in core-collapsed clusters}
\shortauthors{}

\title{White Dwarf Subsystems in Core-Collapsed Globular Clusters}

\correspondingauthor{Kyle Kremer}
\email{kkremer@caltech.edu}

\author[0000-0002-4086-3180]{Kyle Kremer}
\altaffiliation{NSF Astronomy \& Astrophysics Postdoctoral Fellow}
\affiliation{TAPIR, California Institute of Technology, Pasadena, CA 91125, USA}
\affiliation{The Observatories of the Carnegie Institution for Science, Pasadena, CA 91101, USA}

\author[0000-0002-1884-3992]{Nicholas Z. Rui}
\affiliation{TAPIR, California Institute of Technology, Pasadena, CA 91125, USA}

\author[0000-0002-9660-9085]{Newlin C. Weatherford}
\affil{Department of Physics \& Astronomy, Northwestern University, Evanston and Center for Interdisciplinary Exploration \& Research in Astrophysics (CIERA), IL 60208, USA}

\author[0000-0002-3680-2684]{Sourav Chatterjee}
\affil{Tata Institute of Fundamental Research, Homi Bhabha Road, Mumbai 400005, India}

\author[0000-0002-7330-027X]{Giacomo Fragione}
\affil{Department of Physics \& Astronomy, Northwestern University, Evanston and Center for Interdisciplinary Exploration \& Research in Astrophysics (CIERA), IL 60208, USA}

\author[0000-0002-7132-418X]{Frederic A. Rasio}
\affil{Department of Physics \& Astronomy, Northwestern University, Evanston and Center for Interdisciplinary Exploration \& Research in Astrophysics (CIERA), IL 60208, USA}

\author[0000-0003-4175-8881]{Carl L. Rodriguez}
\affil{McWilliams Center for Cosmology, Department of Physics, Carnegie Mellon University, Pittsburgh, PA 15213, USA}

\author[0000-0001-9582-881X]{Claire S. Ye}
\affil{Department of Physics \& Astronomy, Northwestern University, Evanston and Center for Interdisciplinary Exploration \& Research in Astrophysics (CIERA), IL 60208, USA}

\begin{abstract}
Numerical and observational evidence suggests that massive white dwarfs dominate the innermost regions of core-collapsed globular clusters by both number and total mass. Using NGC 6397 as a test case, we constrain the features of white dwarf populations in core-collapsed clusters, both at present day and throughout their lifetimes. The dynamics of these white dwarf subsystems have a number of astrophysical implications. We demonstrate that the collapse of globular cluster cores is ultimately halted by the dynamical burning of white dwarf binaries. We predict core-collapsed clusters in the local universe yield a white dwarf merger rate of $\mathcal{O}(10\rm{)\,Gpc}^{-3}\,\rm{yr}^{-1}$, roughly $0.1-1\%$ of the observed Type Ia supernova rate. We show that prior to merger, inspiraling white dwarf binaries will be observable as gravitational wave sources at milli- and decihertz frequencies. Over $90\%$ of these mergers have a total mass greater than the Chandrasekhar limit. If the merger/collision remnants are not destroyed completely in an explosive transient, we argue the remnants may be observed in core-collapsed clusters as either young neutron stars/pulsars/magnetars (in the event of accretion-induced collapse) or as young massive white dwarfs offset from the standard white dwarf cooling sequence. Finally, we show collisions between white dwarfs and main sequence stars, which may be detectable as bright transients, occur at a rate of $\mathcal{O}(100\rm{)\,Gpc}^{-3}\,\rm{yr}^{-1}$ in the local universe. We find that these collisions lead to depletion of blue straggler stars and main sequence star binaries in the centers of core-collapsed clusters.
\vspace{1cm}
\end{abstract}

\section{Introduction}
\label{sec:intro}

For a standard initial mass function, present-day globular clusters are expected to harbour tens of thousands of white dwarfs (WDs) formed from the evolution of stars with initial masses in the range $\approx1-8\,M_{\odot}$  \citep[e.g.,][]{Salpeter1955,Kroupa2001}. Despite their high abundances, the observational study of WDs in globular clusters has progressed rather slowly, with a number of early searches impeded simply by the low luminosities of these objects \citep[e.g.,][]{Richer1978,OrtolaniRosino1987,Elson1995}. In the mid-1990s, large populations of WDs and clear evidence of the WD cooling sequence were observed for the first time in several nearby globular clusters: M4 \citep{Richer1995,Richer1997}, NGC 6752 \citep{Renzini1996}, and NGC 6397 \citet{Cool1996}. More recently, several studies have taken further strides forward, using measurements of WD cooling sequences to constrain the ages of several Galactic globular clusters \citep[e.g.,][]{Hansen2002,Hansen2007,Hansen2013}. 

The presence of WDs in globular clusters has a number of implications across astronomy. If accreting material from a binary companion, WDs may be observed as cataclysmic variables \citep[e.g.,][]{Grindlay1995,Ivanova2006,Belloni2016,Kremer2020}, which are well-observed in GCs \citep[e.g.,][]{Knigge2012}. WDs may be a critical ingredient in the formation of pulsars \citep[also well-observed in GCs;][]{Ransom2008} through accretion-induced collapse \citep[e.g.,][]{Bailyn1990,Tauris2013,Ye2018}. Additionally, WD mergers facilitated by dynamical interactions in the dense cores of GCs may provide an avenue for Type Ia supernovae or other transients \citep[e.g.,][]{SharaHurley2002,Shen2019,Kremer2020}.

Recent studies have shown that in the central regions of core-collapsed GCs, massive WDs are likely the dominant stellar population \citep[e.g.,][]{Kremer2020,Rui2021b}. The process of cluster core-collapse, in turn, is expected to be facilitated primarily by stellar-mass black holes \citep[BHs; e.g.,][]{Kremer2019d}. Thus, the dynamical evolution of massive WDs
is intimately related to the dynamics governing stellar-mass BHs and a discussion of WDs in GCs merits a discussion of their more massive remnant relatives.

The evidence for stellar-mass BHs in dense star systems like GCs is mounting. Over the past decade, eight stellar BH candidate binaries have been identified in Milky Way GCs through radio/X-ray observations \citep[five candidates;][]{Strader2012,Chomiuk2013,Miller-Jones2015,Shishkovsky2018} and radial-velocity measurements \citep[three candidates;][]{Giesers2018,Giesers2019}. On the computational side, a number of recent studies have shown that populations of BHs play an important role in the dynamical evolution of their host clusters. Once BHs mass-segregate to their host cluster's core, energy generated through series of binary-mediated BH encounters effectively heats the core of the cluster, delaying the onset of cluster core collapse until the BH population is sufficiently depleted \citep[e.g.,][]{Merritt2004,Mackey2007,BreenHeggie2013,Wang2016,ArcaSedda2018,Kremer2018b,Kremer2019a,AntoniniGieles2020}. In some scenarios, this ``BH burning'' may be prominent enough to accelerate the dissolution of a cluster through its tidal boundary \citep[e.g.,][]{Chatterjee2017a,Giersz2019,Weatherford2021,Gieles2021}. Broadly speaking, by leveraging this understanding of the interplay of BHs with their host clusters' structure, it is possible to constrain BH populations indirectly through analysis of observed cluster properties \citep[e.g.,][]{Weatherford2018,Weatherford2020,Askar2018, ArcaSedda2018}.

Recently, \citet{Vitral2021} demonstrated that the nearby core-collapsed NGC 6397 contains a diffuse dark inner subcluster of unresolved objects with a total mass of roughly $1000\,M_{\odot}$ concentrated within the innermost 6 arcsec. NGC 6397 has long been associated with similar claims; a number of analyses have argued this cluster hosts an intermediate-mass BH \citep[IMBH; e.g.,][]{Kamann2016}. However, claims of IMBHs in core-collapsed clusters are controversial  \cite[e.g.,][]{heggie2007core,mcnamara2003does,kirsten2012no,gieles2018mass,tremou2018maveric}. While allowing for the possibility of WDs and neutron stars, \citet{Vitral2021} suggest that stellar-mass BHs likely dominate the mass of this central dark population. However, this is inconsistent with the current consensus understanding of BH dynamics in dense star clusters \citep{Rui2021b}. To this point, several recent papers have shown specifically that, on the basis of observed features, NGC 6397 likely harbours at most a handful of stellar BHs at present day \citep{Weatherford2020,Rui2021}.

\begin{deluxetable*}{l|ccc|c|cccc|ccc}
\tabletypesize{\footnotesize}
\tablewidth{0pt}
\tablecaption{List of $N$-body models \label{table:models}}
\tablehead{
	\colhead{} &
	\colhead{} &
	\colhead{} &
	\colhead{} &
	\colhead{} &
	\multicolumn{4}{c}{Number of White Dwarfs} &
	\multicolumn{3}{c}{Number of WD+WD mergers}\\
	\colhead{$^1$} &
	\colhead{$^2 N\,(\times10^5)$} &
	\colhead{$^3 r_v\,(\rm{pc})$} &
	\colhead{$^4 R_{\rm{gc}}\,(\rm{kpc})$} &
	\colhead{$^5 N_{\rm{BH}}$} &
	\colhead{$^6 N_{\rm{tot}}$} &
	\colhead{$^7 N_{\rm{He}}$} &
	\colhead{$^8 N_{\rm{CO}}$}&
	\colhead{$^{9} N_{\rm{ONe}}$}&
	\colhead{$^{10} N_{\rm{ej}}$}&
	\colhead{$^{11} N_{\rm{in-cluster}}$}&
	\colhead{$^{12} N_{\rm{coll}}$}
}
\startdata
\texttt{1} & 3.9 & 0.95 & 6 & 0 & 36243 & 293 & 34690 & 1260 & 10 & 26 & 31 \\
\texttt{2} & 3.9 & 0.95 & 8 & 0 & 38297 & 355 & 36586 & 1356 & 9 & 18 & 22 \\
\texttt{3} & 3.9 & 1.0 & 6 & 0 & 37364 & 297 & 35828 & 1239 & 11 & 17 & 35 \\
\texttt{4} & 3.9 & 1.0 & 8 & 1 & 37325 & 306 & 35738 & 1281 & 13 & 20 & 21 \\
\texttt{5} & 3.9 & 1.05 & 6 & 1 & 33984 & 313 & 32313 & 1358 & 7 & 8 & 14 \\
\texttt{6} & 3.9 & 1.05 & 8 & 0 & 39302 & 412 & 37471 & 1419 & 3 & 11 & 26 \\
\texttt{7} & 4.0 & 0.95 & 6 & 1 & 37148 & 306 & 35603 & 1239 & 21 & 19 & 30 \\
\texttt{8} & 4.0 & 0.95 & 8 & 0 & 37755 & 244 & 36286 & 1225 & 15 & 16 & 25 \\
\texttt{9} & 4.0 & 1.0 & 6 & 1 & 36067 & 289 & 34404 & 1374 & 1 & 9 & 16 \\
\texttt{10} & 4.0 & 1.0 & 8 & 1 & 37196 & 263 & 35640 & 1293 & 10 & 23 & 28 \\
\texttt{11} & 4.0 & 1.05 & 6 & 0 & 39292 & 359 & 37478 & 1455 & 4 & 10 & 11 \\
\texttt{12} & 4.0 & 1.05 & 8 & 0 & 38223 & 349 & 36443 & 1431 & 2 & 16 & 16 \\
\texttt{13} & 4.1 & 0.95 & 6 & 0 & 39501 & 377 & 37746 & 1378 & 9 & 23 & 26 \\
\texttt{14} & 4.1 & 0.95 & 8 & 0 & 40955 & 410 & 39115 & 1430 & 7 & 12 & 20 \\
\texttt{15} & 4.1 & 1.0 & 6 & 1 & 39257 & 368 & 37515 & 1374 & 11 & 25 & 20 \\
\texttt{16} & 4.1 & 1.0 & 8 & 0 & 40062 & 391 & 38310 & 1361 & 12 & 27 & 28 \\
\texttt{17} & 4.1 & 1.05 & 6 & 0 & 38774 & 394 & 37043 & 1337 & 13 & 15 & 22 \\
\texttt{18} & 4.1 & 1.05 & 8 & 2 & 40439 & 418 & 38529 & 1492 & 3 & 10 & 12 \\
\hline
\hline
\texttt{19} & 4 & 2 & 8 & 9 & 41776 & 499 & 40340 & 937 & 0 & 0 & 0 \\
\texttt{20} & 4 & 4 & 8 & 44 & 40700 & 1054 & 38717 & 929 & 0 & 0 & 0 \\
\hline
\hline
\texttt{21}$^{\alpha}$ & 3.2 & 1 & 6 & 2 & 27260 & 2516 & 23730 & 1014 & 23 & 17 & 33 \\
\enddata
\tablecomments{Complete list of models used in this study. Models \texttt{1-18} are all similar in properties to NGC 6397 at present-day (see Figure \ref{fig:models}). Unlike the other models, models \texttt{19-20} still retain BHs at present and thus have yet to undergo core collapse (see Figure \ref{fig:params}). Model \texttt{21} (denoted by ${\alpha}$) has a primordial binary fraction of $50\%$, while all other models assume a $5\%$ primordial binary fraction (see discussion in Section \ref{sec:primordial_binaries}).}
\end{deluxetable*}

In this paper, we follow up on the initial results presented in \citet{Rui2021b} and demonstrate that the observations presented in \citet{Vitral2021} can be explained by a diffuse subcluster of massive WDs. As the population of stellar BHs is slowly depleted over the lifetime of a cluster, massive WDs become the most massive population of stars.\footnote{Of course, neutron stars are of comparable mass to the most massive WDs. However, the majority of neutron stars are expected to be ejected at early times from clusters due to natal kicks \citep[in a typical cluster, only $1000$ NSs are retained at late times compared to roughly $10^4$ WDs; e.g.,][]{Ye2018}. Thus, WDs are the dominant population for the purposes of this study.} Once nearly all BHs have been ejected and the energy produced through BH dynamics is no longer able to delay collapse of the cluster core, these massive WDs are expected to dominate (by number and mass) the inner regions of their host cluster \citep[e.g.,][]{Kremer2020}. To demonstrate these processes, we present a large set of $N$-body models representative of NGC 6397. These models build upon the analysis of \citet{Rui2021}, which presented best-fit models for NGC 6397 and other clusters (on the basis of matches to the observed surface brightness and velocity dispersion profiles) from an extensive suite of models covering the parameter space of Milky Way GCs \citep[the \texttt{CMC Cluster Catalog};][]{Kremer2020}. We go on to examine broadly populations of WDs in clusters and discuss various physical processes made possible by the high WD densities expected in core-collapsed clusters like NGC 6397.

This paper is organized as follows. In Section \ref{sec:methods}, we describe the $N$-body models used in this study and discuss the specific application to NGC 6397 observations. In Section \ref{sec:results}, we discuss the results from our models and argue that NGC 6397 hosts a dense population of WDs at present. If indeed dense populations of WDs are common in core-collapsed clusters, this implies an amplified rate of WD mergers which, in turn, has important implications for Type Ia supernovae and other possible transients and also the formation of milli- and decihertz gravitational wave (GW) sources. In Section \ref{sec:WD_mergers}, we discuss the dynamics of WD subsystems and estimate the rate of WD mergers and detection prospects for GW detectors.
In Section \ref{sec:outcomes}, we explore possible outcomes of WD mergers/collisions and discuss a few observational tests of these outcomes. In Section \ref{sec:WDMS_coll}, we discuss interactions of WDs with main sequence stars in core-collapsed clusters. Finally, in Section \ref{sec:conclusion}, we summarize our results and discuss several avenues for future study.

\section{Computational methods}
\label{sec:methods}

\subsection{Prescriptions for white dwarf physics}
\begin{figure}
    \includegraphics[width=\columnwidth]{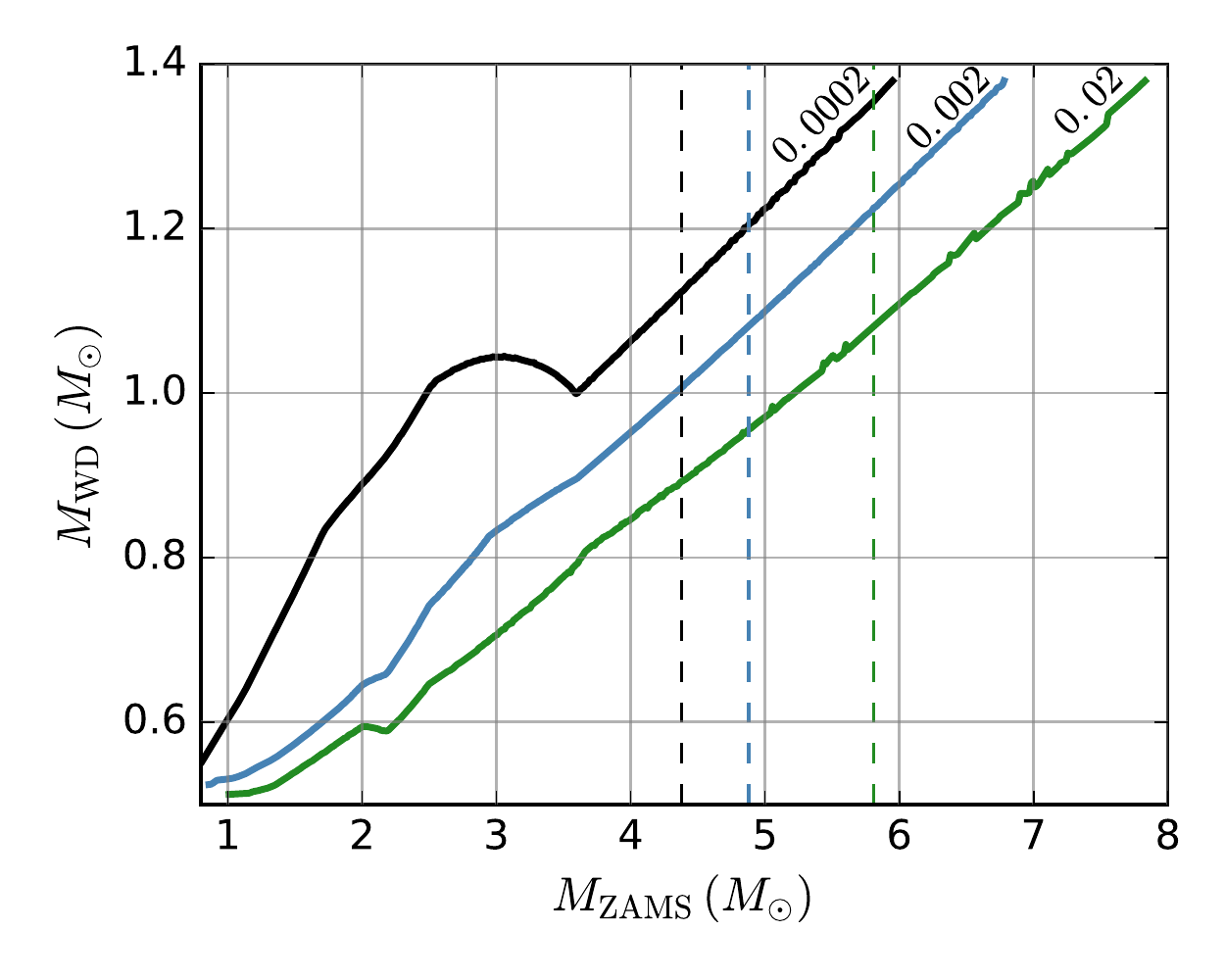}
    \caption{\footnotesize WD--ZAMS mass relation for three different metallicities: $Z=0.0002$ (corresponding to NGC 6397) in black, 0.002 in blue, and 0.02 in green. We show as vertical dashed lines the ZAMS mass corresponding to the transition from CO to ONe WD formation (with colors corresponding to the given metallicity).}
    \label{fig:WD_vs_ZAMS}
\end{figure}

To create our $N$-body cluster models, we use our cluster dynamics code \texttt{Cluster Monte Carlo} (\texttt{CMC}). For a recent review of \texttt{CMC}, including a discussion of the most current prescriptions for compact object formation, see \citet{Kremer2020}. We summarize here the treatments for WD formation and evolution implemented in \texttt{CMC}. Unless otherwise noted, this treatment follows exactly that implemented in \texttt{SSE} \citep{Hurley2000} and \texttt{BSE} \citep{Hurley2002}. We specify three distinct types of WDs:
\begin{enumerate}
    \item \textbf{He WDs:} Helium (He) WDs (\texttt{BSE} type $k=10$) are formed following envelope loss of giant branch stars with zero-age main sequence mass ($M_{\rm{ZAMS}}$) less than the maximum initial stellar mass such that He ignites degenerately in a helium flash \citep[see Equation 2 of][]{Hurley2000}. Note that the envelope stripping required to produce He WDs can only occur through binary evolution. We assume dynamical stellar collisions are all ``sticky sphere,'' meaning that giant envelopes are not ejected \citep[for further detail, see][]{Kremer2020c}.
    \item \textbf{CO WDs:} WDs composed primarily of carbon (C) and oxygen (O) (\texttt{BSE} type $k=11$) are formed by envelope loss of a thermally pulsing asymptotic giant branch (AGB) star with mass below the minimum mass required to undergo non-degenerate C ignition in the core \citep[denoted as $M_{\rm{up}}$ in][]{Pols1998,Hurley2000}. Note that $M_{\rm{up}}$ depends on metallicity \citep[see Table 2 of][]{Pols1998}. By default in \texttt{CMC}, we set $M_{\rm{up}}=1.4\,M_{\odot}$ \citep[see][]{Breivik2020}. At the metallicity of NGC 6397 ($Z\approx0.0002$), the most massive CO WD that will form through single star evolution is roughly $1.1\,M_{\odot}$.
    \item \textbf{ONe WDs:} WDs composed primarily of O and neon (Ne) (\texttt{BSE} type $k=12$) form by envelope loss of a thermally pulsing AGB star with mass in the range $M_{\rm{up}}$ to $M_{\rm{ec}}$, where $M_{\rm{ec}}$ \citep[by default,  $M_{\rm{ec}}=2.5\,M_{\odot}$ in our models; see][]{Breivik2020} is the minimum mass of a star that avoids electron capture onto Ne and Mg in the core \citep[also metallicity dependent; see][]{Pols1998}. For NGC 6397-like metallicity, ONe WDs formed through stellar evolution will have masses $>1.1\,M_{\odot}$.
\end{enumerate}
In Figure \ref{fig:WD_vs_ZAMS} we show the WD mass to zero-age main sequence (ZAMS) mass relation for three different metallicities. The vertical dashed lines indicate the ZAMS mass corresponding to the transition from CO to ONe WD formation as described above (with colors corresponding to the given metallicity).

We compute white dwarf radii as in \citet{Tout1997,Hurley2000}:

\begin{equation}
    \label{eq:radius}
    R_{\rm{WD}} = \max \Bigg[ R_{\rm{NS}}, 0.0115 \sqrt{ \Big( \frac{M_{\rm{Ch}}}{M_{\rm{WD}}} \Big)^{2/3} - \Big( \frac{M_{\rm{WD}}}{M_{\rm{Ch}}} \Big)^{2/3} } \Bigg]
\end{equation}
(in units of $R_{\odot}$), where $R_{\rm{NS}}=10\,$km is the assumed neutron star radius and $M_{\rm{Ch}}=1.44\,M_{\odot}$ is the Chandrasekhar mass.

We model WD luminosity evolution using standard cooling theory \citep{Mestel1952,Hurley2000}:
\begin{equation}
    \label{eq:lum}
    L_{\rm{WD}} = 133 \Bigg(\frac{M_{\rm{WD}}}{M_{\odot}}\Bigg) \Bigg(\frac{Z}{Z_{\odot}}\Bigg)^{0.4}
    \Bigg(A\Big[\frac{t}{\rm{Myr}}+0.1\Big]\Bigg)^{-1.4} L_{\odot},
\end{equation}
where $Z$ is the metallicity, $t$ is the age of the WD since formation, and $A$ is the effective baryon number, which is dependent on WD composition. We use $A=4$ for He WDs, $A=15$ for CO WDs, and $A=17$ for ONe WDs. If a new WD is formed through a WD--WD merger, we assume the age is reset to zero (see Section \ref{sec:outcomes}).

Of particular relevance to this study are the outcomes of dynamically-mediated collisions between WDs and dynamical mass transfer onto WDs (i.e., when a WD binary is driven to Roche lobe overflow, and ultimately merger, through GW-driven inspiral). We describe below our methods to compute the masses and types of remnants resulting from these interactions.

Mass transfer occurs in a double WD binary when it has inspiralled sufficiently through GW emission that one of the WDs fills its Roche lobe. Upon Roche lobe overflow, we assume dynamically unstable mass transfer occurs (leading to a merger) when the mass ratio of the system $q > 0.628$ \citep{Tout1997}. Stable mass transfer ensues for $q < 0.628$, possibly leading to a long-lived accreting binary \citep[for a description of the treatments implemented in \texttt{BSE} for stable mass transfer from one WD onto another, see Section 2.6.6 of][]{Hurley2002}. The outcome of dynamically unstable mass transfer depends on the types of WDs involved. We assume that when two He WDs merge, the triple-$\alpha$ reaction ignites and releases sufficient energy to completely unbind the merger product, leaving behind no remnant \citep{Hurley2002}. When a He WD accretes dynamically onto a CO/ONe WD, we assume the accreted material swells to form an envelope around the CO/ONe core and forms a He giant branch star \citep[following][]{Iben1996}. The most relevant case for the purposes of this study is dynamical accretion of material from one CO/ONe WD onto another CO/ONe WD. In this case, the more massive WD is assumed to accrete total mass $\Delta M=\dot{M}_{\rm{Edd}}\tau_{\rm{\dot{M}}}$ from the donor star, where $\dot{M}_{\rm{Edd}}$ is the Eddington limit and $\tau_{\rm{\dot{M}}} = \sqrt{\tau_{\rm{KH}}\tau_{\rm{dyn}}}$ is the characteristic mass-transfer timescale (here, $\tau_{\rm{KH}}$ and $\tau_{\rm{dyn}}$ are the Kelvin-Helmholtz and dynamical timescales of the donor, respectively). For typical WDs of mass roughly $1\,M_{\odot}$, $\dot{M}_{\rm{Edd}}\approx 10^{-5}M_{\odot}\rm{yr}^{-1}$ and $\tau_{\rm{\dot{M}}}\approx2000\,$yr, giving a typical $\Delta M \approx 10^{-3}\,M_{\odot}$. Following \citet{SiaoNomoto1998,Belczynski2002}, if $M_1+\Delta M > M_{\rm{Ch}}$, we assume electron capture leads to an accretion-induced collapse of the primary, leaving behind a neutron star of mass $M=1.24\,M_{\odot}$. If $M_1+\Delta M < M_{\rm{Ch}}$ (the more common case), we assume an ONe/CO WD primary remains (depending on the accretor's composition). In both cases, the secondary (donor) is destroyed.

A dynamically-mediated collision occurs when two WDs pass one another such that the pericenter distance $r_p<R_1+R_2$, the summed radii of the two components. When this occurs, we adopt the ``sticky sphere'' approximation that the collision product's mass $M_3$=$M_1$+$M_2$, the summed component masses. The stellar type of the collision product depends on the types of the two colliding objects \citep[for further detail, see Section 2.7 of][]{Hurley2002}. For collisions between two He WDs, we assume the triple-$\alpha$ reaction is ignited so that the collision remnant is destroyed completely. For collisions between a He WD and a CO/ONe WD, we assume the He material forms an envelope around the CO/ONe core such that a giant branch He star forms (\texttt{BSE} stellar type $k=8\,\rm{or}\,9$) with core mass equal to the original CO/ONe WD mass. For collisions between two CO and/or ONe WDs, we assume a new CO/ONe WD is formed unless $M_3>M_{\rm{Ch}}$, in which case we assume a NS forms through accretion-induced collapse.

Of course, these treatments of merger/collision products are highly simplified and general outcomes of WD mergers are likely more complex and diverse than considered here \citep[e.g.,][]{IbenTutukov1984,Webbink1984,Nomoto1991,HillebrandtNiemeyer2000,WoosleyKasen2011,Shen2012,Schwab2016,Schwab2021}. However, we stress that from the perspective of cluster dynamics, the specific outcome is largely irrelevant. Only a few dozen WD mergers/collisions are expected to occur in a typical core-collapsed cluster (see Table \ref{table:models}), so merger/collision products (or lack thereof for an explosive transient outcome) will constitute at most a few percent of the total population of $\sim10^3$ WDs in the innermost region and $\sim0.1\%$ of the total population of $\gtrsim10^4$ WDs in the entire cluster. However, if detectable, the outcomes of these few dozen WD mergers may potentially provide powerful observational tests of the WD dynamics described in this paper. We return to this topic in more detail in Section \ref{sec:outcomes}.

Finally, when a MS star of mass $M_1$ merges with a WD of mass $M_2$, we assume the MS star is accreted dynamically and swells up to form an envelope around the WD, which becomes the core of a new giant star with total mass $M_3 = M_1+M_2$ and core mass $M_{3,c}=M_2$. For a He WD+MS merger, the product will be on the first giant branch star (\texttt{BSE} stellar type $k=3$). For a CO/ONe WD+MS merger, the product is a thermally pulsating AGB star (\texttt{BSE} stellar type $k=6$). A new age appropriate to its core mass is then computed for this new star as described in Section 2.7.4 of \citet{Hurley2002}. This treatment is also simplified in nature. In reality, a collision between a MS star and a WD may lead to complete destruction of the star and detonation of the WD \citep[e.g.,][]{SharaShaviv1977}. We return to this topic in Section \ref{sec:WDMS_coll}.

\begin{figure}
    \includegraphics[width=\columnwidth]{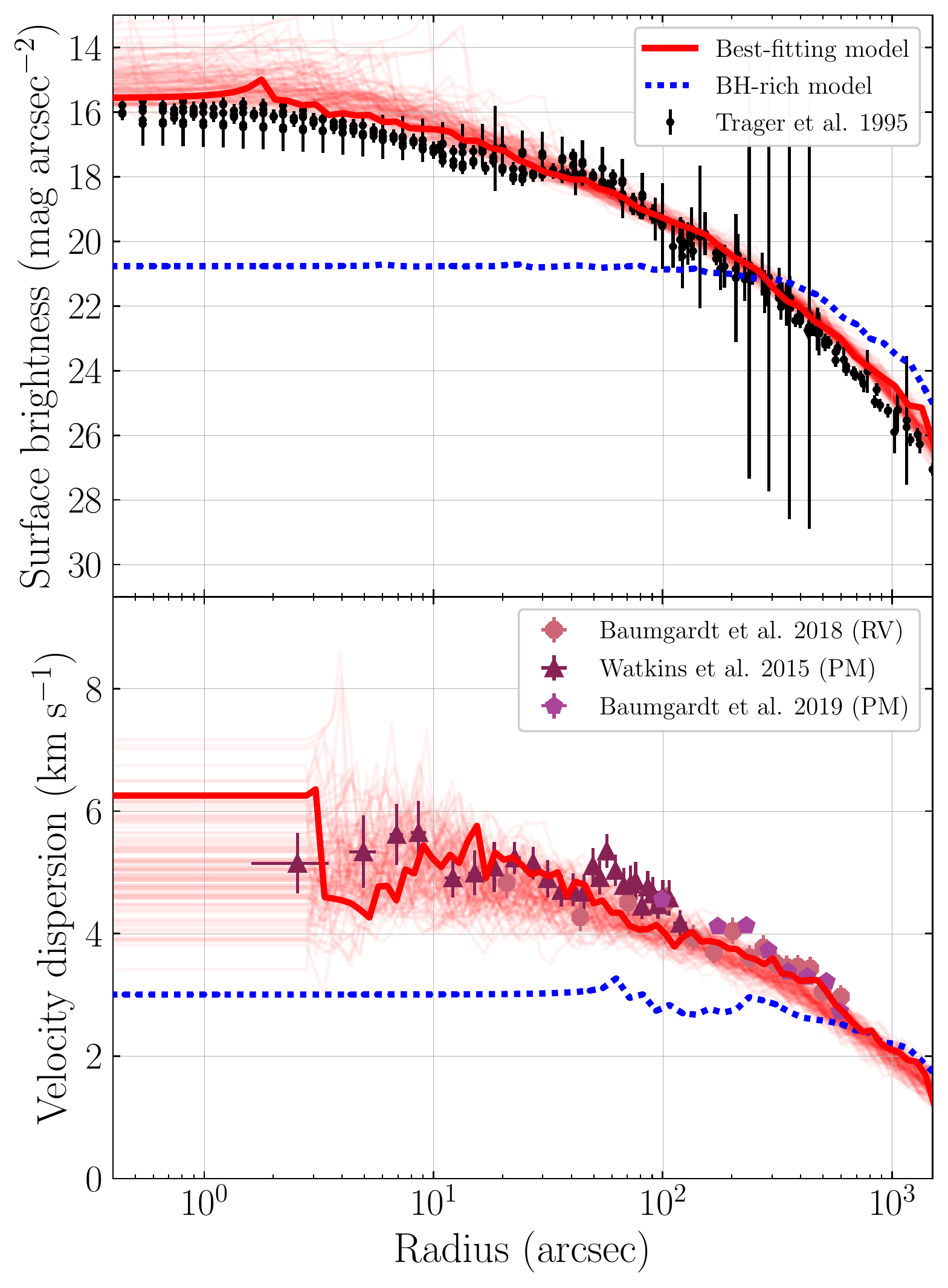}
    \caption{\footnotesize Surface brightness (top) and velocity dispersion (bottom) profiles for all models in this study compared to observed profiles for NGC 6397. In red, we show data from simulations \texttt{1-18}, with the bold red curves indicating our best-fit model for NGC 6397. In blue we show results from simulation \texttt{20} which, unlike the other models, contains a BH subsystem throughout its entire lifetime.}
    \label{fig:models}
\end{figure}

\subsection{Cluster models}

In \citet{Rui2021}, we presented a framework for comparing \texttt{CMC} models to observed clusters based on surface brightness and velocity dispersion profiles. As an application of this framework, we analyzed the \texttt{CMC Cluster Catalog} and identified best fit models for dozens of GCs observed in the MW, including NGC 6397, which was examined in detail in this earlier study. We direct the reader to \citet{Rui2021} for a more detailed discussion of our scheme for identifying best fit models for a given observed system.

In this analysis, we start with the \texttt{CMC} simulation identified by \citet{Rui2021} as an effective match to present-day NGC 6397. The initial conditions for this simulation are shown as row \texttt{10} in Table \ref{table:models}. In order to increase our sample of NGC 6397-like models, we run 17 additional simulations that explore the parameter space near our original best-fit model. Specifically, we adopt a grid of values with initial particle number $N/10^5=(3.9,4,4.1)$,
initial virial radius $r_v/\rm{pc}=(0.95,1,1.05)$, and Galactocentric distances $R_{\rm{gc}}/\rm{kpc}=(6,8)$. We also include two additional simulations with $r_v=2$ and $4\,$pc to demonstrate the differences between an NGC 6397-like cluster and BH-rich clusters with large cores at present-day. 

In all models, we adopt a metallicity of $Z=0.0002$ \citep[the observed metallicity of NGC 6397;][]{Harris1996}. We assume all models are initially fit by a King model \citep{King1962} with concentration parameter $W_0=5$. We adopt a Kroupa initial mass function in the range $0.08-150\,M_{\odot}$ \citep{Kroupa2001}. For primordial binaries, we assume orbital separations drawn from a log-uniform distribution in orbital period from near contact to the hard-soft boundary, mass ratios drawn from a uniform distribution from $0.1-1$, and eccentricities drawn from a thermal distribution. For our fiducial set of models (\texttt{1-20} in the table), we adopt a primordial binary fraction of $5\%$. We also examine a simulation with a primordial binary fraction of $50\%$ to explore its effect on our results (see Section \ref{sec:primordial_binaries}).

In Figure \ref{fig:models}, we show surface brightness and velocity dispersion profiles for all models in this study overlaid on observed profiles of NGC 6397 from \citet{Trager1995,Watkins2015,Baumgardt2018,Baumgardt2019}. For further discussion of how models are compared to observations, see \citet{Rui2021}. In bold red, we show the model identified as the best fit to NGC 6397 \citep[via the method in][]{Rui2021}, corresponding to simulation \texttt{17} from Table \ref{table:models} at the final cluster time snapshot. In dashed blue, we show simulation \texttt{20} at the final cluster snapshot. Although nearly identical in total mass to our NGC 6397-like models, model \texttt{20}, which contains $44$ BHs, clearly does not resemble the overall structure of NGC 6397 at present, as its core is influenced significantly by its BH population. In light red, we show all remaining NGC 6397-like models from Table \ref{table:models} (models \texttt{1-18}).

\bigskip

\section{White dwarf populations in core-collapsed clusters}
\label{sec:results}

\begin{figure}
    \includegraphics[width=\columnwidth]{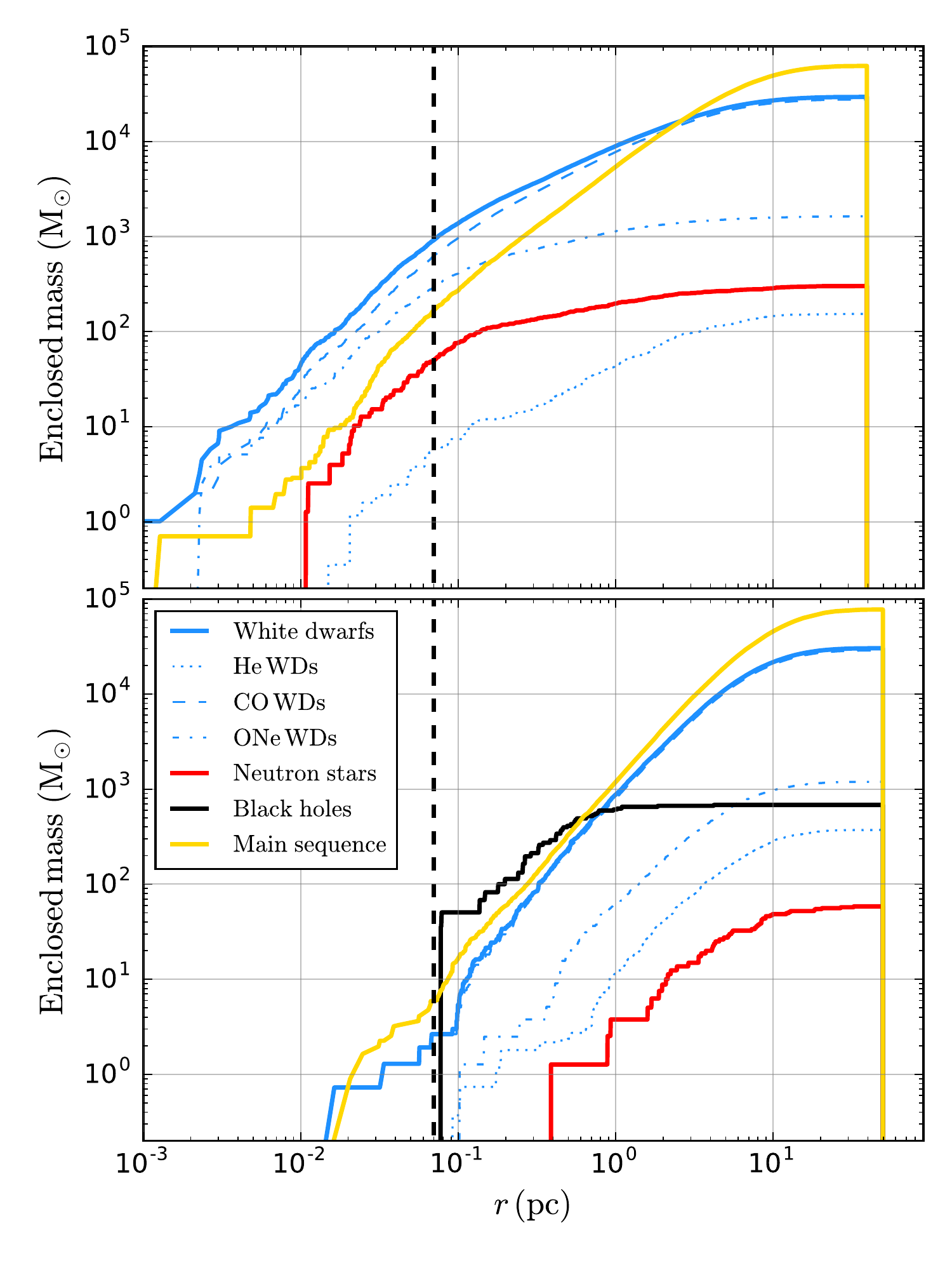}
    \caption{\footnotesize Enclosed mass versus radius for various stellar populations for our best-fit model for NGC 6397 (simulation \texttt{17}; top panel) and for our representative BH-rich cluster (simulation \texttt{20}; bottom panel). The vertical dashed black line denotes the cluster-centric distance within which \citet{Vitral2021} constrained a diffuse subsystem of roughly $1000\,M_{\odot}$ in NGC 6397. We argue that, in NGC 6397, WDs are the most plausible explanation for this massive dark population.
    }
    \label{fig:radial_dist}
\end{figure}

In Figure \ref{fig:radial_dist}, we show the enclosed mass of various stellar populations as a function of (2D-projected) distance from the cluster center for two representative clusters: in the top panel we show our best-fit model to NGC 6397 (model \texttt{17} in Table \ref{table:models}) and in the bottom panel we show the BH-rich counterpart model \texttt{20} that has yet to undergo core collapse by the present day. In both panels, the blue curves denote the WD population: all WDs (solid), CO WDs (dashed), He WDs (dotted), and ONe WDs (dash-dotted). The remaining colors denote other stellar populations, as shown in the figure legend.

We denote with a vertical dashed line a distance of $0.07\,$pc ($6\,$arcsec) from the cluster center, within which \citet{Vitral2021} argue a diffuse cluster of dark objects of mass $1000-2000\,M_{\odot}$ is enclosed in NGC 6397. As shown, roughly $10^3\,M_{\odot}$ ($30\,M_{\odot}$) of WDs (NSs) are enclosed within this distance in our NGC 6397-like model. This model contains zero BHs. Thus, as in \citet{Rui2021b}, we argue that a central population of massive WDs -- roughly two-thirds and one-third CO and ONe WDs, respectively -- is the most plausible explanation for the dark diffuse component identified in \citet{Vitral2021}.

\begin{figure}
    \includegraphics[width=\columnwidth]{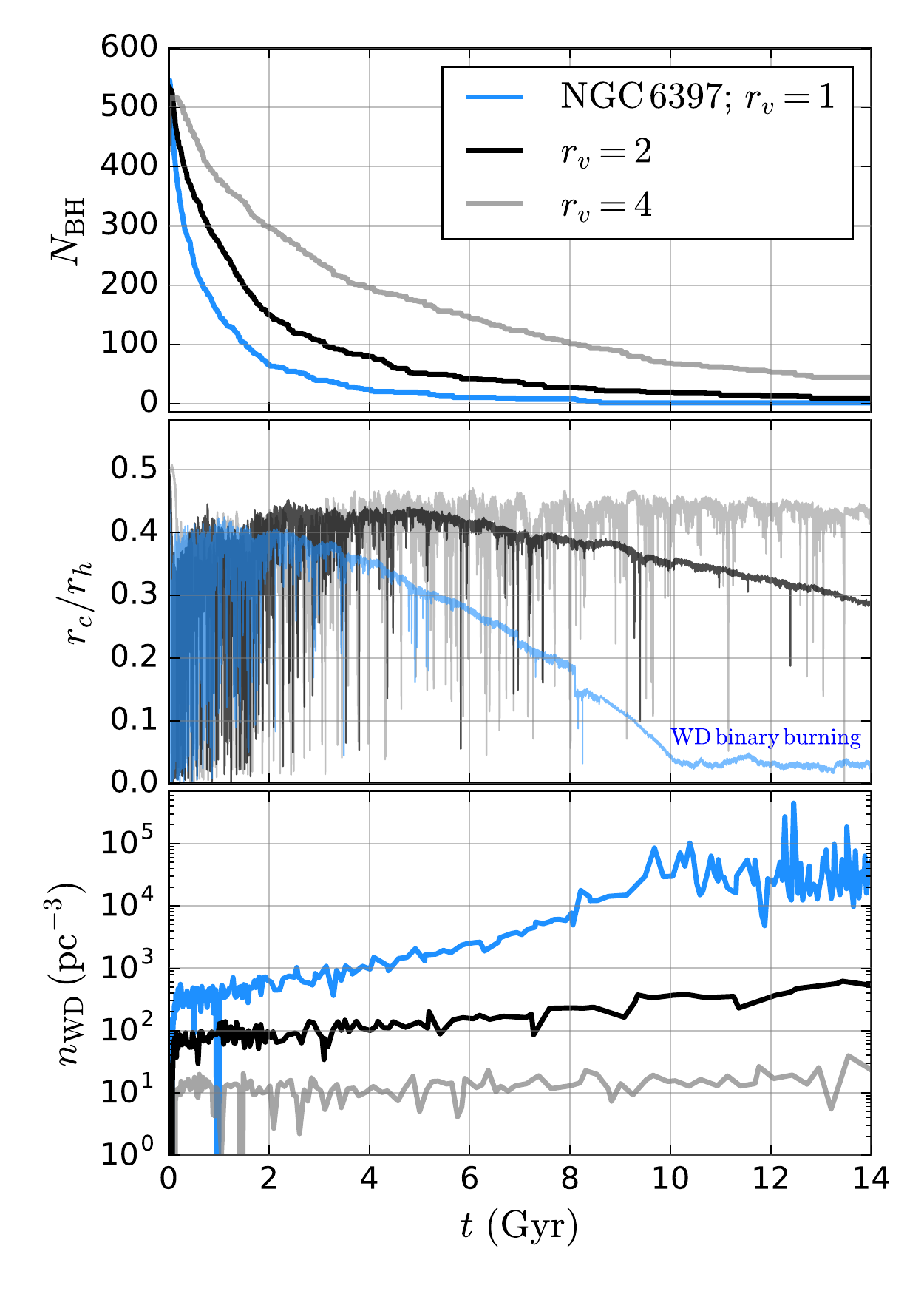}
    \caption{\footnotesize The evolution in time of various cluster properties for an NGC 6397-like model (blue; $r_v=1\,$pc) and two non-core-collapsed BH-rich clusters of identical initial mass: black ($r_v=2\,$pc) and gray ($r_v=4\,$pc). In the top panel we show the total number of retained BHs versus time, in the middle we show the ratio of core to half-light radius, $r_c/r_h$, and in the bottom panel we show the number density of WDs within the core radius. In our NGC 6397-like model, core contraction is halted at $t\approx10\,$Gyr when quasi-equilibrium is attained through energy generated by dynamical ``burning'' of WD binaries.}
    \label{fig:params}
\end{figure}

The model shown in the bottom panel of Figure \ref{fig:radial_dist} contains roughly $1000\,M_{\odot}$ worth of stellar BHs ($N_{\rm{BH}}=44$) at late times. A BH population this massive dominates the core dynamics and mass segregation forces lower-mass luminous stellar populations outward. As a result, this BH-rich counterpart exhibits a large observed core radius (roughly $3\,$pc; see Figure \ref{fig:params}) at present. The key point is that a large BH population is inconsistent with an observed core-collapsed structure.

In Figure \ref{fig:params}, we show the time evolution of these same two cluster models (blue curves denote our best-fit simulation for NGC 6397 while gray curves show the BH-rich simulation \texttt{20}). We also include simulation \texttt{19} as an intermediary ($r_v=2\,$pc) between these two extremes. The top panel shows the number of BHs retained as the cluster evolves. In all three clusters, roughly 500 BHs (out of 800 total) are retained in the cluster at early times after taking into account natal kicks. The number of BHs subsequently declines as BHs are ejected due to dynamical processes as described in Section \ref{sec:intro}. The number of BHs retained at late times is determined by the ejection rate, which in turn is determined by the initial cluster density; the NGC 6397-like model (initial $r_v=1\,$pc) is initially the most dense of the three and ultimately ejects all of its BHs.

A frequently used parameter that reflects the dynamical age of a GC is the ratio between the core radius, $r_c$, and the half-mass radius, $r_h$ \citep[e.g.,][]{Fregeau2007,Chatterjee2013a}. In a self-gravitating system like a GC, energy flows naturally from the core to the halo leading to core contraction and halo expansion \citep[e.g.,][]{HeggieHut2003}. As a cluster ages, $r_c/r_h$ decreases \citep[at a rate determined primarily by ``BH burning'' in the cluster core;][]{Kremer2019d}. This core contraction halts when quasi-equilibrium is achieved through the energy generated by dynamical burning of stellar binaries \citep[e.g.,][]{HeggieHut2003}. In the middle panel of Figure \ref{fig:params}, we show $r_c/r_h$ versus time. Here the core radius is defined as the theoretical density-weighted core radius typically used in $N$-body simulations \citep[e.g.,][]{CasertanoHut1985}. In all three simulations, the core expands at early times due to mass loss from high-mass stellar evolution. In the $r_v=4\,$pc model which retains the most BHs at the end of the simulation, the core is heated by BH dynamics and $r_c/r_h$ remains roughly constant over the complete lifetime. For $r_v=2\,$pc, the core is still contracting by the end of the simulation---binary burning equilibrium has yet to be achieved. In the NGC 6397-like cluster, which ejects all of its BHs after roughly $10\,$Gyr, the core contracts significantly, ultimately undergoing collapse and attaining quasi-equilibrium once binary burning (primarily of WDs, as shown in Figure \ref{fig:radial_dist}) begins.

Finally, in the bottom panel, we show the number density of WDs within the cluster core radius \citep[here defined in the observational sense; see][]{Morscher2015} versus time. The number of retained WDs is determined primarily by the initial mass function and subsequent stellar evolution, which are assumed to be identical for these three clusters. Thus, the total number of WDs is roughly comparable in these three clusters (see also Table \ref{table:models}). In this case, the differences in the WD density evolution are attributed primarily to the core radius evolution. We see that, in the absence of BHs, a core-collapsed cluster like NGC 6397 features a central WD density roughly three orders-of-magnitude higher than a cluster in the BH burning phase.

\begin{figure}
    \includegraphics[width=\columnwidth]{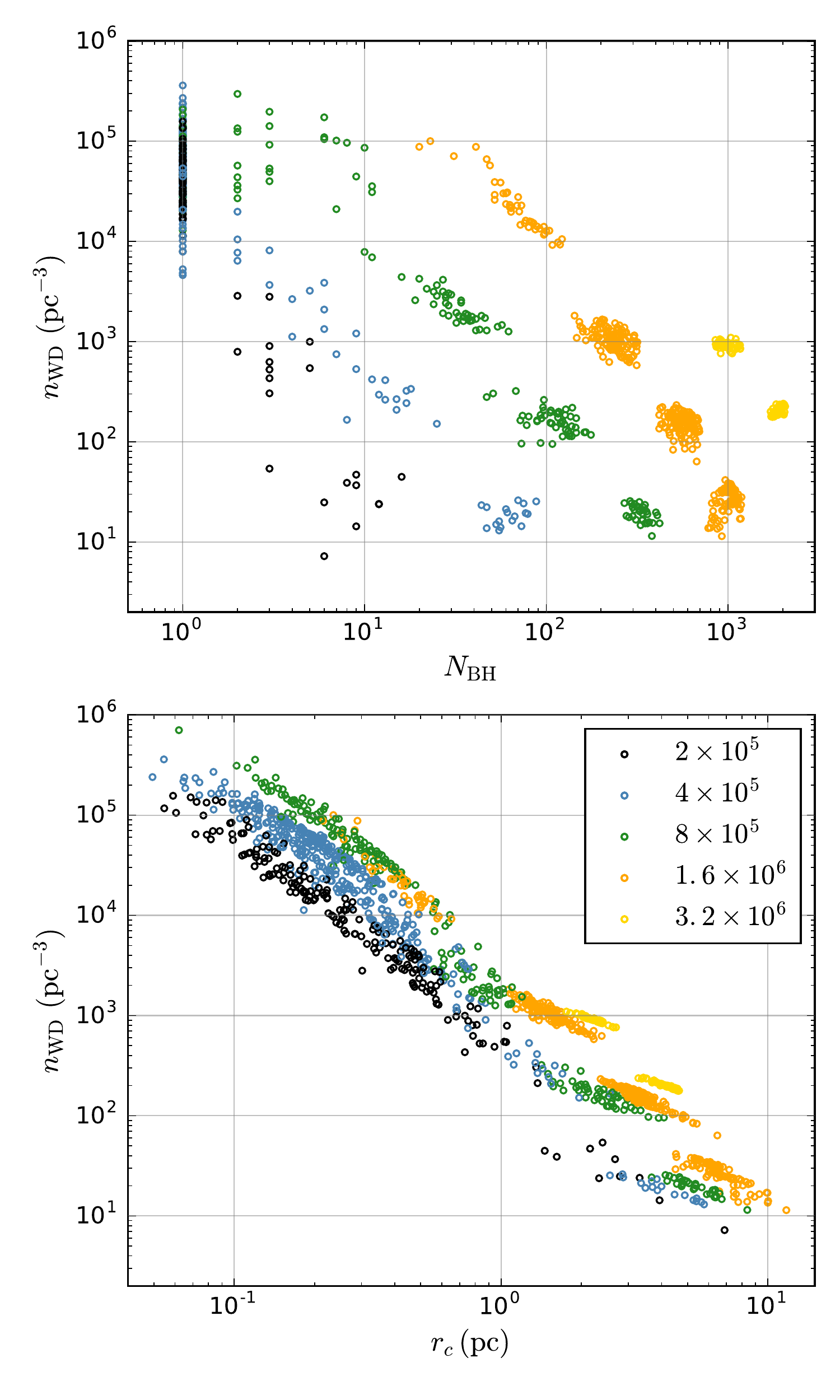}
    \caption{\footnotesize Number density of WDs within the cluster core radius versus total number of BHs retained (top panel) and versus core radius (bottom panel) for all low-metallicity cluster models in the \texttt{CMC Cluster Catalog} \citep{Kremer2020}. Black, blue, green, orange, and yellow points indicate models with initial $N$ of $[2,4,8,16,32]\times10^5$, respectively.}
    \label{fig:rho_vs_BH}
\end{figure}

The connection between WD and BH subsystems suggested by Figures \ref{fig:radial_dist} and \ref{fig:params} in the context of clusters with NGC 6397-like masses and properties is indeed a general trend across all cluster types. To demonstrate this point, we use the set of cluster models published in the \texttt{CMC Cluster Catalog}. This catalog contains roughly 150 cluster simulations (evolved for a Hubble time or until dissolution) that collectively span the range in observed properties (such as cluster mass, metallicity, radii, and Galactocentric distances) of the Milky Way GCs at present day. To focus specifically on old low-metallicity clusters\footnote{While in principle WD subsystems will develop in high-metallicity (e.g., near $Z_{\odot}$) clusters, high-metallicity clusters born relatively recently are unlike to be sufficiently old at present-day to have reached core-collapse (although of course this depends on cluster masses and other properties). We reserve for future work a study of the possibility of WD subsystems in high-metallicity young clusters.}, we exclude here the 50 solar metallicity simulations in the catalog. In Figure \ref{fig:rho_vs_BH}, we show the number density of WDs within the cluster core radius versus total number of retained BHs (top panel) and versus the core radius (bottom panel) for all models in this catalog with ages in the range $10-14\,$Gyr. The different colors indicate the five distinct values of initial $N$ (initial cluster mass) assumed. As illustrated in the figure, for constant initial cluster mass, clusters with larger BH subsystems (and thus larger core radii due to the influence of BH burning) feature relatively lower density WD populations in their cores. Once its BH subsystem has been depleted, a cluster develops a dense WD subsystem.

Motivated by these results, we argue that a dense central population of WDs is a common feature of all dynamically-old core-collapsed clusters. Indeed, the presence of high density WD subsystems in core-collapsed clusters has been hinted at previously, specifically in the context of M15 \citep{murphy2011fokker}. Given that core-collapsed clusters are common \citep[in the Milky Way, roughly $20\%$ of GCs are core-collapsed;][]{Harris1996}, this has important implications for various astrophysical phenomena that may be influenced by WD dynamics. We go onto to explore several such applications in the following sections.

\section{White dwarf mergers}
\label{sec:WD_mergers}

We expect WD binaries to merge through two primary mechanisms:

\begin{enumerate}
    \item \textit{Gravitational wave-driven inspirals} that occur after a binary has hardened to Roche lobe overflow through series of binary-mediated encounters. If $q>0.628$ (see Section \ref{sec:methods}), we assume a merger ensues. These GW inspiral-driven mergers may occur both inside and outside the cluster, depending on if a binary has attained sufficiently high recoil to have been ejected from its host.
    \item \textit{Direct physical collisions} that occur when the pericenter distance of a pair of WDs is less than the sum of the WD radii, $r_p<R_1+R_2$. These collisions may occur through both single--single encounters or during binary resonant encounters.\footnote{The direct collision channel is analogous to the ``GW capture'' channel for binary BH mergers \citep[e.g.,][]{Samsing2018,Rodriguez2018b}. For WDs, the cross section for GW capture lies within the WD's physical radius, thus in this limit, a head-on collision (i.e., with no inspiral) occurs.}
\end{enumerate}

To develop a sense of the typical timescales at play and to build physical intuition, we analytically explore these merger processes in Section \ref{sec:analytic_rate}. In Section \ref{sec:model_rate}, we compare the results from our $N$-body models to our analytic estimates. In Section \ref{sec:rate}, we estimate the overall WD merger rate and discuss the potential detectability of these binaries as GW sources.

\begin{figure*}
    \includegraphics[width=\linewidth]{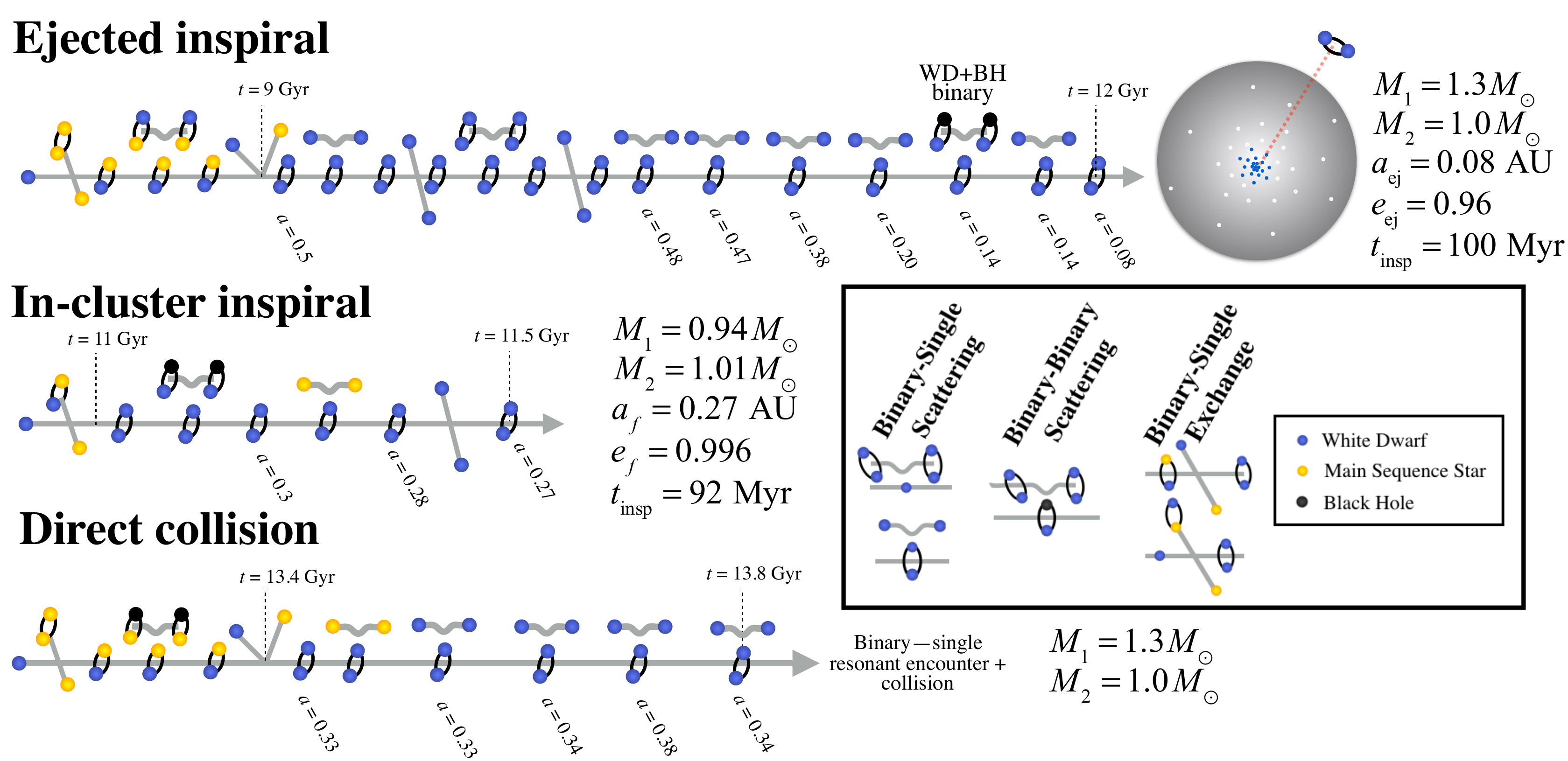}
    \caption{\footnotesize Illustration of the dynamical encounter histories for representative WD mergers from each of the three merger channels described in Section 
    \ref{sec:WD_mergers}. These mergers are each taken from simulation \texttt{17} in Table \ref{table:models}.}
    \label{fig:interactions}
\end{figure*}

\subsection{Analytic estimates}
\label{sec:analytic_rate}

Within a core-collapsed cluster, WD binaries will undergo series of binary-mediated encounters that gradually harden the binaries \citep[e.g.,][]{HeggieHut2003}. Conservation of energy requires when a binary hardens that energy be transferred to the center-of-mass velocity of the interacting stars. Thus, the recoil velocity of the binary increases with each encounter. This process continues until either the binary receives a recoil velocity sufficient to eject it from the host cluster or until the binary's GW inspiral time becomes small enough that the binary merges inside the cluster.

The characteristic orbital separation at which a binary will be dynamically ejected is given by

\begin{equation}
    a_{\rm{ej}} \approx \alpha \frac{G m}{v_{\rm{esc}}^2} \approx 0.02\,\rm{AU} \left( \frac{\it{m}}{\it{M}_{\odot}} \right) \left( \frac{\it{v}_{\rm{esc}}}{40\,km\,s^{-1}} \right)^{-2} 
\end{equation}
\citep[e.g.,][]{Rodriguez2016a, Samsing2018}, where we have adopted $v_{\rm{esc}}\approx 40 \,\rm{km\,s}^{-1}$ as the escape velocity from the core of a host cluster representative of NGC 6397 and assumed $\alpha \approx 1/6 \times [1/\delta-1]$, where $\delta\approx7/9$ is a constant that reflects the amount the semi-major axis of a binary changes with each encounter \citep[for detail, see][]{Samsing2018}. To determine whether or not a binary with orbital separation $a_{\rm{ej}}$ is more likely to be ejected (and potentially merge after ejection) or merge inside its host cluster, we compare the characteristic timescales for GW inspiral and binary--single encounters that may lead to ejection.

For a binary near $a_{\rm{ej}}$, the characteristic timescale to undergo a binary--single encounter and be ejected is given by

\begin{equation}
    t_{\rm{ej}} \approx \frac{1}{n_{\rm{WD}} \Sigma \sigma_v},
\end{equation}
where $n_{\mathrm{WD}}$ is the number density of single WDs in the WD subsystem, $\sigma_v$ is the velocity dispersion, and $\Sigma$ is the cross section of the binary. Including gravitational focusing, this cross section can be expressed as

\begin{equation}
    \label{eq:sigma_bs}
    \Sigma = \pi a_{\rm{ej}}^2 \left(1 + \frac{6Gm}{a_{\rm{ej}} \sigma_v^2} \right),
\end{equation}
where $m\approx1\,M_{\odot}$ is a typical WD mass. Assuming the gravitational focusing term dominates (typical for velocity dispersions found in GCs), we obtain 

\begin{multline}
    \label{eq:t_ej}
    t_{\rm{ej}} \approx 900\,\rm{Myr} \left( \frac{\sigma_v}{10\,\rm{km\,s}^{-1}} \right)  \\ 
    \times \left( \frac{v_{\rm{esc}}}{40\,\rm{km\,s}^{-1}} \right)^2 \left( \frac{\it{n}_{\rm{WD}}}{10^6\,\rm{pc}^{-3}} \right)^{-1} \left( \frac{\it{m}}{\it{M}_{\odot}} \right)^{-2}
\end{multline}
for binaries near $a_{\rm{ej}}$.

For highly eccentric binaries \citep[representative of dynamically-formed binaries; e.g.,][]{HeggieHut2003}, the GW inspiral time for a WD binary at this separation can be estimated \citep[as in][]{Peters1964} as

\begin{equation}
    \label{eq:t_insp}
    t_{\rm{insp}} \approx 100\,\rm{Myr} \left( \frac{\it{a}}{0.02\,\rm{AU}} \right)^4 \left(\frac{\it{m}}{\it{M}_{\odot}} \right)^{-3} \left(1-e^2\right)^{7/2},
\end{equation}
where we have adopted a fiducial eccentricity of $e=0.9$.

With these two timescales in mind, we can draw two immediate conclusions: First, given that both $t_{\rm{ej}}$ and $t_{\rm{insp}}$ are small compared to the typical time we expect an NGC 6397-like cluster to have reached a core-collapsed configuration with high central WD density ($\approx4\,$Gyr; see Figure \ref{fig:params}), we conclude that GW inspirals of WDs will certainly occur in these environments. Second, since $t_{\rm{insp}} < t_{\rm{ej}}$, we expect that once a binary is hardened to $a_{\rm{ej}}$, there is a higher probability of an in-cluster merger relative to an ejection.

Note that this is different from the case of binary BHs. Since $a_{\rm{ej}}\propto m$ and since $m_{\rm{BH}} \sim 10\,M_{\odot}$ is an order of magnitude larger than the typical WD mass, the characteristic orbital separation for BH binaries to be ejected is relatively wide compared to WD binaries. In this contrasting case for BHs, $t_{\rm{ej}} \propto m^{-2} \approx 9\,$Myr (Equation \ref{eq:t_ej}), a factor of 100 times shorter than the characteristic ejection time for WD binaries. Meanwhile, the GW inspiral time of a binary of separation $a_{\rm{ej}}$ scales as $t_{\rm{insp}}\propto a_{\rm{ej}}^4\, m^{-3} \propto m$. Thus for BH binaries, $t_{\rm{insp}}$ is an order of magnitude or more larger than $t_{\rm{ej}}$, indicating that most BH binaries will be ejected prior to merger. Indeed, in the models considered in this study, roughly two-thirds of all BH mergers occur after the binary has been ejected with the other third occurring through in-cluster inspirals and GW captures combined. Of course, the ratio of in-cluster to ejected mergers depends upon the host cluster's escape velocity; more massive and compact clusters will always feature a higher ratio of in-cluster to ejected mergers, regardless of the binary component masses. We have focused here simply on escape velocities representative of NGC 6397.

\begin{figure*}
    \includegraphics[width=\linewidth]{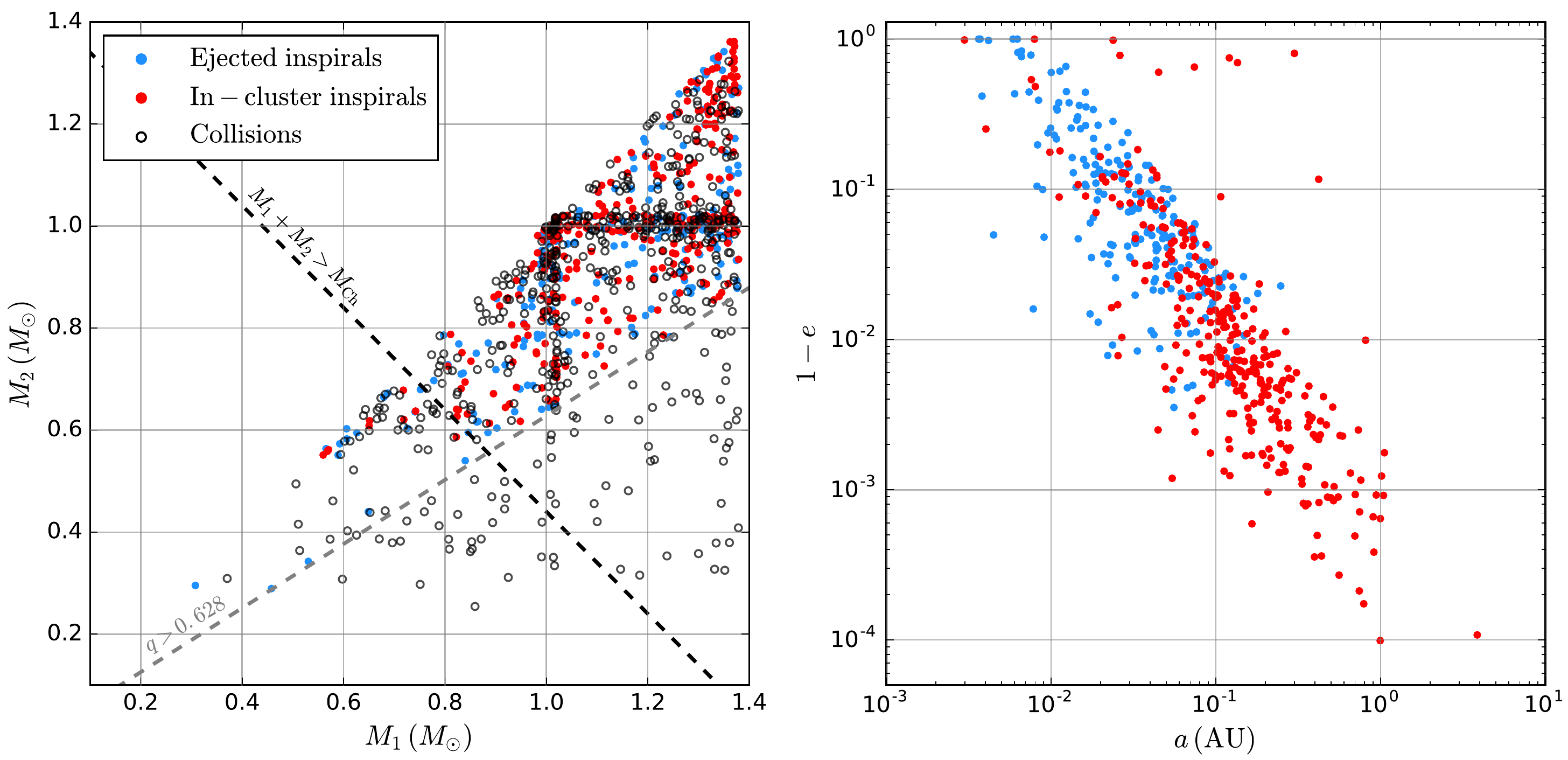}
    \caption{\footnotesize \textit{Left panel:} Component masses for all WD--WD mergers/collisions occurring at late times $t>9\,$Gyr) in our simulations. Blue and red circles denote ejected and in-cluster inspirals, respectively, and black open circles denote collisions. The dashed black line marks systems with a total mass equal to the Chandrasekhar limit ($1.44\,M_{\odot}$) and the dashed gray curve marks the $q>0.628$ boundary above which Roche lobe overflow is assumed to lead to merger (see Section \ref{sec:methods}). \textit{Right panel:} Eccentricity versus semi-major axis at time of last dynamical encounter for all inspirals, with colors the same as the left panel.}
    \label{fig:masses}
\end{figure*}

Next, we estimate the rate of WD mergers occurring through direct collisions. For single--single encounters, this rate is given by

\begin{equation}
    \Gamma_{\rm{coll}}^{\rm{ss}} \approx
    n_{\rm{WD}}\Sigma_{\rm{coll}} \sigma_v N_{\rm{WD}};
\end{equation}
where $\Sigma_{\rm{coll}}$ is the cross section for collisions (Equation \ref{eq:sigma_bs} with $a$ replaced by the WD's physical radius). This produces the following scaling:
    
\begin{multline}
    \label{eq:coll_ss}
    \Gamma_{\rm{coll}}^{\rm{ss}} \approx 0.2\,\rm{Gyr}^{-1}
    \left( \frac{\it{m}}{\it{M}_{\odot}} \right)
    \left( \frac{\it{R}_{\rm{WD}}}{10^8\,\rm{cm}} \right) \\
    \times \left(\frac{\it{n}_{\rm{WD}}}{10^6\,\rm{pc}^{-3}} \right)
    \left( \frac{\sigma_v}{10\,\rm{km\,s}^{-1}} \right)^{-1}
    \left( \frac{\it{N}_{\rm{WD}}}{1000} \right).
\end{multline}
For a cluster spending roughly $4\,$Gyr in a core-collapsed configuration (as in the NGC 6397-like model shown in Figures \ref{fig:radial_dist} and \ref{fig:params}), we can expect $\mathcal{O}(1)$ WD--WD collisions through single--single encounters.

For collisions occurring during binary--single resonant encounters, the collision rate can be expressed as 

\begin{equation}
    \Gamma_{\rm{coll}}^{\rm{bs}} \approx n_{\rm{WD}}\Sigma \sigma_v N_{\rm{bin}} P_{\rm{coll}},
\end{equation}
where $\Sigma$ is the cross section for binary--single encounters given by Equation \ref{eq:sigma_bs}, $N_{\rm{bin}}$ is the number of WD binaries (assuming a $10\%$ binary fraction in the core, $N_{\rm{bin}} \approx 0.1N_{\rm{WD}}$), and $P_{\rm{coll}}$ is the probability that a given binary resonant encounter leads to a collision. For resonant encounters with roughly equal component masses, $P_{\rm{coll}} \propto R/a$ \citep[e.g.,][]{HeggieHut2003}. For example, \citet{Samsing2017} showed that for binary--single encounters with three WDs and semi-major axis values of $10^{-3}-0.1\,$AU, $P_{\rm{coll}} \sim 100(R/a)$. Adopting this relation and, as before, assuming that gravitational focusing dominates, we write the binary--single collision rate as

\begin{multline}
    \label{eq:coll_bs}
    \Gamma_{\rm{coll}}^{\rm{bs}} \approx 3\,\rm{Gyr}^{-1}
    \left( \frac{\it{m}}{\it{M}_{\odot}} \right)
    \left( \frac{\it{R}_{\rm{WD}}}{10^8\,\rm{cm}} \right) \\
    \times \left(\frac{\it{n}_{\rm{WD}}}{10^6\,\rm{pc}^{-3}} \right)
    \left( \frac{\sigma_v}{10\,\rm{km\,s}^{-1}} \right)^{-1}
    \left( \frac{\it{N}_{\rm{WD}}}{1000} \right).
\end{multline}
Therefore, we expect collisions through binary--single encounters to be roughly a factor of ten more common compared to those occurring through single--single interactions. For NGC 6397-like clusters, we can expect $\mathcal{O}(12)$ binary-mediated collisions per cluster. Since the characteristic timescale for collisions $t_{\rm{coll}} \approx (\Gamma_{\rm{coll}}^{\rm{bs}})^{-1} \approx 300\,$Myr is comparable to the timescale for in-cluster inspirals (Equation \ref{eq:t_insp}) we can expect these two channels to contribute roughly comparable rates in NGC 6397-like clusters.

\subsection{Results from N-body models}
\label{sec:model_rate}

Having motivated the topic analytically, we now discuss the WD mergers identified in our $N$-body models. In total, from our 18 models representative of NGC 6397 (see columns 10, 11, and 12 of Table \ref{table:models}), we identify 161, 305, and 403 mergers from the ejected inspiral, in-cluster inspiral, and head-on collision channels, respectively. As predicted from the analytic estimates, in-cluster inspirals and collisions produce comparable rates in our models, while the rate from ejected inspirals is lower by a small factor. Of the $403$ total direct collisions, 15 occur through single--single encounters and 196 (192) occur through binary--single (binary--binary) resonant encounters; as predicted from the analytic estimates, binary-mediated encounters dominate the direct collision rate by an order of magnitude or more.

In Figure \ref{fig:interactions}, we illustrate the sequence of dynamical encounters leading to merger for three representative mergers, one from each channel. All three of these sequences come from simulation \texttt{17}. As demonstrated by these three examples, WD binaries in general undergo extended series of encounters with other WDs and MS stars in their host clusters, hardening the binaries until GW inspiral leads to merger. En route to a sufficiently compact orbital separation, WDs will occasionally collide with other WDs (and MS stars; see Section \ref{sec:WDMS_coll}).

In the left panel of Figure \ref{fig:masses}, we show the component masses for all WD mergers and collisions identified in our models. More than $90\%$ of the mergers/collisions feature a total mass in excess of the Chandrasekhar limit (shown as the dashed black line in the figure). As a dashed gray line, we show the $q>0.628$ boundary, above which WD binaries driven to Roche lobe overflow through GW inspiral are assumed to be dynamically unstable, leading to merger (see Section \ref{sec:methods}). In the right panel of Figure \ref{fig:masses}, we show the semi-major axis and eccentricity at the time of the last dynamical encounter for all WD binaries that merge through GW inspiral. For in-cluster (ejected) mergers, the median semi-major axis is $0.13\,$AU ($0.03\,$AU) and the median eccentricity is $0.99$ ($0.94$). As expected for reasons described in Section \ref{sec:analytic_rate}, in-cluster mergers typically form at higher eccentricity and higher semi-major axis compared to ejected mergers.

\begin{figure*}
    \begin{center}
    \includegraphics[width=0.7\linewidth]{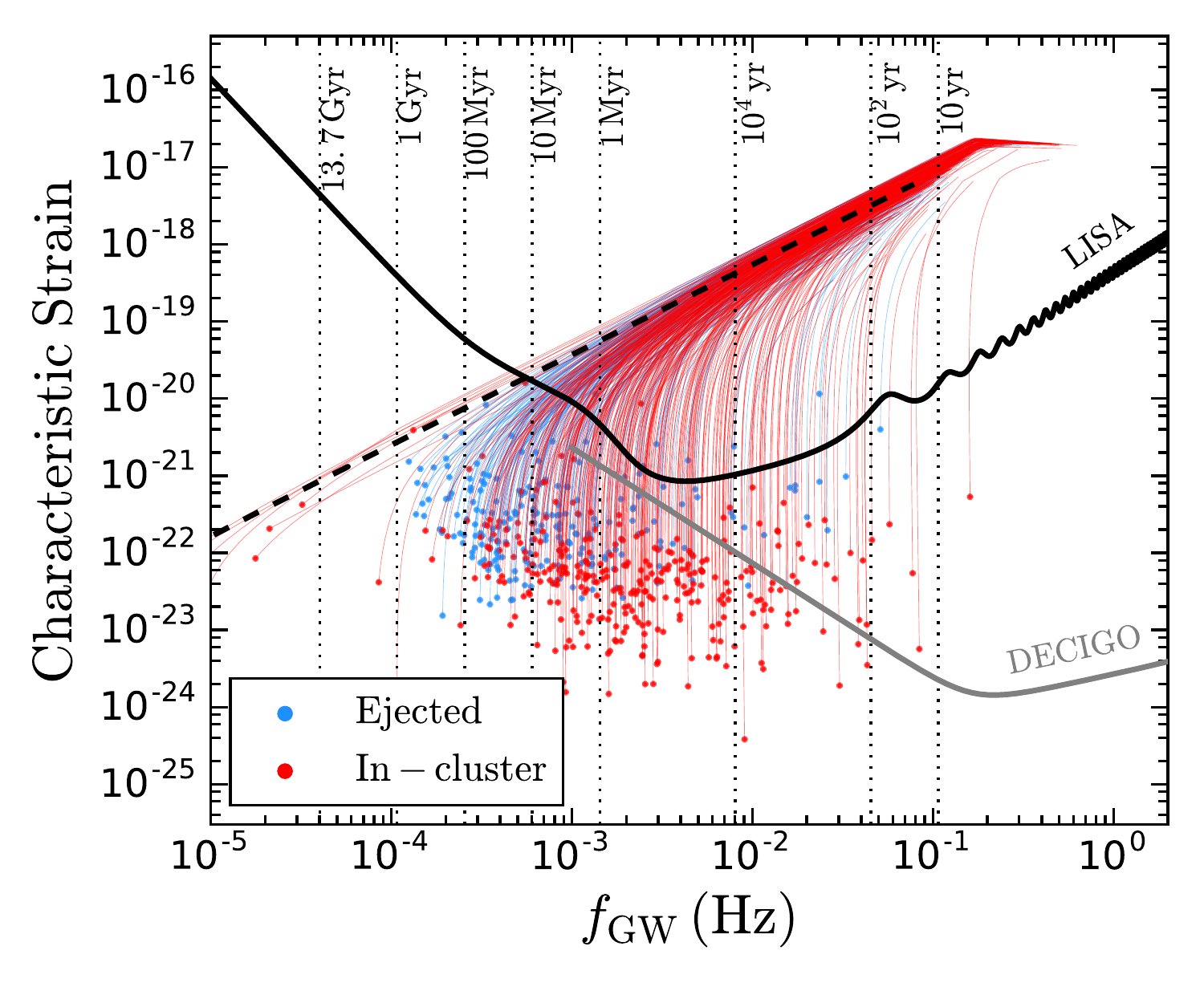}
    \caption{\footnotesize Characteristic strain versus peak GW frequency for all WD--WD inspirals identified in our models. Red (blue) curves mark systems that merge inside (outside) their host cluster. For reference, the dashed black curve denotes a representative $1\,M_{\odot}+1\,M_{\odot}$ system and the vertical dashed lines mark inspiral times for this representative binary as it evolves. The solid black and gray curves show the sensitivity curves for LISA and DECIGO, respectively. \label{fig:LISA} }
    \end{center}
\end{figure*}

\subsection{Estimating the merger rate and prospects for GW detection}
\label{sec:rate}

Assuming a typical core-collapsed cluster underwent core collapse and formed a WD subsystem roughly $4\,$Gyr ago (as motivated by the NGC 6397-like models; see Figure \ref{fig:params}), we estimate from our models a total rate of roughly $10^{-8}$ WD mergers per year for a single NGC 6397-like cluster. In the Milky Way, roughly $20\%$ of observed GCs are core-collapsed. Assuming a population of roughly $200$ GCs in the Milky Way (150 observed plus an additional $\sim50$ that may be hidden behind the bulge), we assume roughly 40 core-collapsed clusters populate our Galaxy at present. Accounting for the fact that NGC 6397 is a factor of roughly two times less massive than the average Milky Way GC, and assuming that the WD merger rate scales roughly linearly with cluster mass \citep[as is the case for binary BH mergers; e.g.,][]{Rodriguez2016a}, we estimate a total WD merger rate of roughly $10^{-6}\,\rm{yr}^{-1}$ in a Milky Way-like galaxy. For a giant elliptical galaxy like M87, expected to host $\gtrsim10^4$ GCs \citep[e.g.,][]{Tamura2006}, the rate may be as high as $10^{-4}\,\rm{yr}^{-1}$. Assuming the average number density of GCs in the local universe is roughly $3\,\rm{Mpc}^{-3}$ \citep[e.g.,][]{Rodriguez2016a} and assuming a fraction $f$ of these clusters are core-collapsed, we estimate an overall WD merger rate of roughly $f \times50\,\rm{Gpc}^{-3}\,\rm{yr}^{-1}$ in the local universe from GCs.

In analogy with the proposed connection of BH subsystems in GCs and gravitational-wave sources detected by LIGO/Virgo \citep[e.g.,][]{Rodriguez2016a,Askar2017,Fragione2018b,RodriguezLoeb2018,Kremer2020,Antonini2019}, an obvious application to explore is GW astronomy. Unlike stellar BH and NS binaries which merge within the LIGO/Virgo frequency band ($f_{\rm{GW}} > 10\,$Hz), WD binaries merge at decihertz frequencies, as limited by the physical sizes of WDs. In this case, millihertz detectors such as LISA \citep[e.g.,][]{Amaro-Seoane2017} and decihertz detectors like DECIGO \citep[e.g.,][]{Kawamura2011,ArcaSedda2020} are necessary to observe the inspiral and merger of these systems.

In Figure \ref{fig:LISA}, we show the characteristic GW strain \citep[calculated as in][assuming a LISA observation time of 5 years]{Kremer2019b} and peak frequency of GW emission \citep[calculated as in][]{Wen2003} from formation (marked as the filled circles) to merger for all inspiraling systems identified our models. In red (blue), we show the WD binaries that merge inside (outside) their host cluster, depending on whether or not a given system was dynamically ejected. For reference, the dashed black curve denotes the inspiral of a characteristic $1\,M_{\odot}+1\,M_{\odot}$ circular WD binary. For this characteristic system, we show a number of representative inspiral times as vertical dashed lines. For all systems in this plot, we assume a source distance of $9\,$kpc from Earth, the average heliocentric distance of GCs in the MW.

The solid black and gray curves denote the sensitivity curves for LISA \citep[][]{Robson2019} and DECIGO. For ``stationary'' sources (i.e., sources whose frequency evolves negligibly over the $5\,$yr LISA observation window), the signal-to-noise ratio can be read roughly by eye from this figure as the height of the characteristic strain above the sensitivity curve. For instance, at $f_{\rm{GW}}\approx1.5\times10^{-3}\,$Hz ($t_{\rm{insp}}\approx1\,$Myr), the characteristic strain of the characteristic $1\,M_{\odot}+1\,M_{\odot}$ binary (dashed black curve) is roughly a factor of 10 above the LISA sensitivity curve at this frequency, indicating a signal-to-noise ratio of roughly 10 at the assumed distance of $9\,$kpc. 

The characteristic strain of a source is inversely proportional to the distance. Thus, WD--WD inspirals are unlikely to be resolved by LISA at distances far beyond the Milky Way. Combined with the small estimated WD merger rate from Galactic GCs ($\approx10^{-6}\,\rm{yr}^{-1}$), this suggests it is highly unlikely that LISA would observe a merging WD binary formed in a GC. However, LISA may resolve these inspiraling binaries as stationary sources at lower frequencies (roughly mHz) when their inspiral times are still long \citep[e.g.,][]{Willems2007,Kremer2018c}. We can estimate the number of detectable stationary sources as $N_{\rm{stat}} \approx \Gamma_{\rm{merger}} \times t_{\rm{insp}}$, where $\Gamma_{\rm{merger}}$ is the total merger rate (roughly $10^{-8}\,\rm{yr}^{-1}$ per typical core-collapsed GC) and $t_{\rm{insp}}$ is the inspiral time when the source becomes resolvable by LISA. From Figure \ref{fig:LISA}, we see $t_{\rm{insp}}\approx10\,$Myr for a $1\,M_{\odot}+1\,M_{\odot}$ binary when its strain exceeds the LISA sensitivity. Thus, we estimate roughly $10^{-1}$ sources per core-collapsed cluster. NGC 6397, roughly a factor of fives times nearer to Earth than the average MW GC, may contain up to order unity resolvable sources. Assuming 40 core-collapsed clusters in the MW, we estimate roughly 4 resolvable sources from GCs in our Galaxy. To within a factor of a few, this estimate is satisfyingly consistent with the roughly 20 resolvable WD--WD sources previously estimated in \citet{Kremer2018c} using \texttt{CMC} models that neither included relativistic effects nor focused specifically on core-collapsed clusters.

Although LISA will not be able to observed extragalactic WD sources in large numbers, third-generation decihertz detectors (which propose to probe higher GW frequencies at greater sensitivity), may observe many extragalactic WD binaries. For instance, \citet{Maselli2020} demonstrated that mergers of roughly $1\,M_{\odot}$ WDs can be resolved by DECIGO with signal-to-noise ratios of roughly 20 out to a distance of $1\,$Gpc, potentially enabling GW detection of roughly dozens of WD mergers formed in GCs per year.

\begin{figure}
    \begin{center}
    \includegraphics[width=\columnwidth]{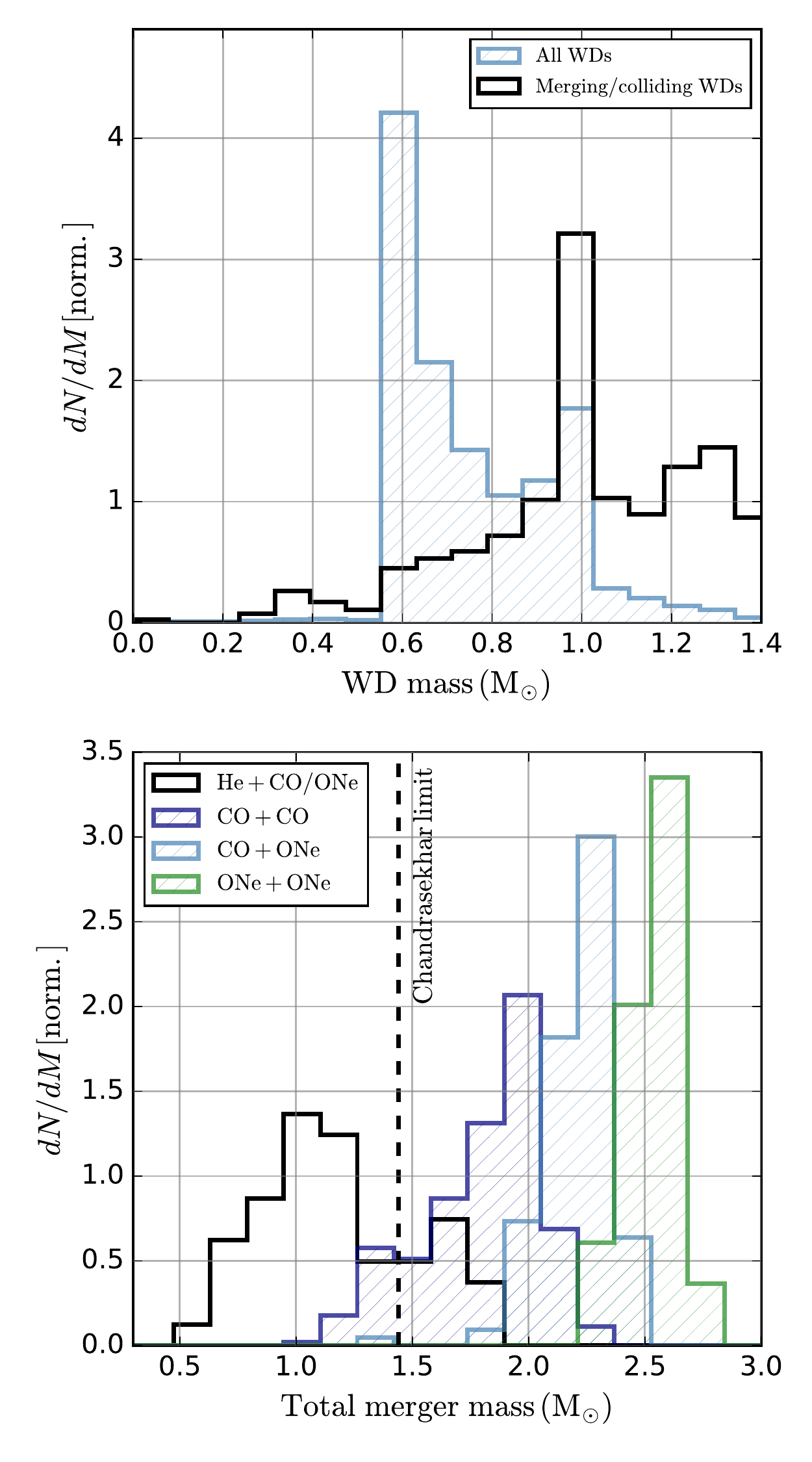}
    \caption{\footnotesize \textit{Top panel:} Normalized mass distribution for all WDs formed through standard single/binary star evolution (light blue) and mass distribution for all WD components that undergo mergers/collisions with other WDs (black). \textit{Bottom panel:} Normalized distribution of total mass, $M_1+M_2$, for all WD mergers/collisions in our models. Different colors denote different combinations of WD types; $44\%$ of all mergers are CO+CO WDs (dark blue histogram), $38\%$ are CO+ONe (light blue), $13\%$ are ONe+ONe (green), and $5\%$ are He+CO/ONe (black). \label{fig:mass_dist_total} }
    \end{center}
\end{figure}

\section{White dwarf merger outcomes}
\label{sec:outcomes}

In this section, we discuss potential outcomes of the WD mergers/collisions identified in our models and examine a few potential observational tests.

In the top panel of Figure \ref{fig:mass_dist_total}, we show in black the normalized distribution of component masses for all WD mergers (here combining the ejected and in-cluster inspirals and the collisions). In blue, we show the normalized mass distribution for all WDs formed through standard stellar evolution. The peak at roughly $0.6\,M_{\odot}$ arises simply from the initial mass function; these WDs come from stars with $M_{\rm{ZAMS}}\approx 1\,M_{\odot}$, which in old ($t\gtrsim10\,$Gyr) clusters are the least massive objects that have yet to form WDs. Meanwhile, the peak in the WD mass distribution at roughly $1\,M_{\odot}$ arises from the ``bump'' in the $M_{\rm{WD}}-M_{\rm{ZAMS}}$ relation at ZAMS masses in the range $\approx2.5-3.5\,M_{\odot}$ at low metallicities (see Figure \ref{fig:WD_vs_ZAMS}). This bump is caused by the growth (or lack thereof) of the progenitor stars' cores while on the giant branch \citep[see][]{Hurley2000}. As shown, the population of WDs that go on to merge are preferentially more massive than the overall WD population. In a core-collapsed cluster, the most massive WDs mass-segregate to the dense central regions where, due to increased dynamical activity, they are more likely to merge.

In the bottom panel of Figure \ref{fig:mass_dist_total}, we show the normalized distribution of total mass, $M_1+M_2$, for all merging/colliding WD pairs of various types identified in our models. Of the roughly $1000$ mergers/collisions identified in our models, $44\%$ feature two CO WDs (dark blue histogram), $13\%$ two ONe WDs (green histogram), $38\%$ a CO and ONe WD pair (light blue histogram), and only $5\%$ a He WD paired with a CO or ONe WD (black histogram). The dashed vertical line marks the Chandrasekhar limit. $91\%$ of all mergers/collisions in our models have a total mass greater than $M_{\rm{Ch}}$.

The outcomes of WD mergers/collisions are highly uncertain and various scenarios have been proposed depending on the masses and compositions of the two WDs involved. These outcomes fall broadly into two main camps: the non-destructive case (similar to what is implemented in \texttt{CMC}) that forms a massive single WD \citep[e.g.,][]{Schwab2021} or NS/pulsar \citep[e.g.,][]{NomotoIben1985,Schwab2016} and the destructive case where an explosive transient results with no remnant left behind. As described in Section \ref{sec:methods}, we assume in \texttt{CMC} that all WD mergers/collisions avoid complete detonation and instead result in an ONe WD (in the event of dynamically unstable binary mass transfer) or an accretion-induced-collapse NS (in the event of a ``sticky sphere'' dynamical collision).

\begin{figure}
    \includegraphics[width=\columnwidth]{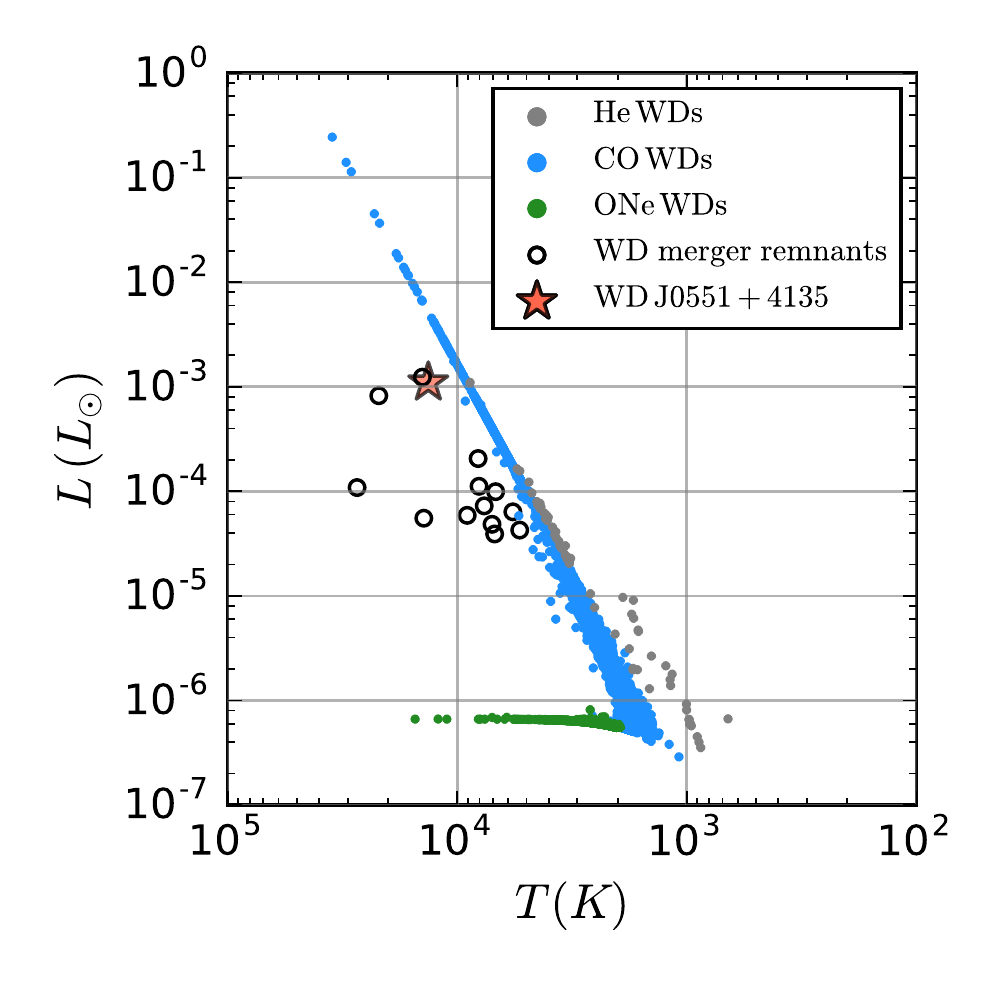}
    \caption{\footnotesize Hertzprung-Russell diagram for all single WDs in the best-fit model identified for NGC 6397. Gray, blue, and green filled circles denote He, CO, and ONe WDs, respectively. Open circles denote young massive WDs formed through recent WD--WD mergers. The red star shows the inferred properties of WD J0551+4135, a high-mass WD candidate in the Galactic field proposed to have formed from a previous WD merger similar to those studied here.
    }
    \label{fig:HR_diag}
\end{figure}

In Figure \ref{fig:HR_diag}, we show a Hertzprung-Russell diagram for all single WDs found in our best-fit model for NGC 6397. Luminosities are calculated as in Equation \ref{eq:lum} and temperatures are computed using the radii obtained from Equation \ref{eq:radius} and assuming blackbody emission. Filled circles denote WDs formed through standard single star evolution. The blue band reproduces the well-known cooling track produced by CO WDs of various ages \citep[e.g.,][]{Hansen2007,Richer2013}. ONe WDs result from a relatively narrow range in ZAMS mass ($\approx4.5-6\,M_{\odot}$ for $Z=0.0002$; see Section \ref{sec:methods}). So, unlike the CO WDs that form over a wide range of times, ONe WDs all form at $t\sim100\,\rm{Myr}$. This relatively narrow range in ONe WD age manifests itself as the horizontal green band seen in the figure.

Open circles show remnant WDs formed through recent in-cluster WD mergers identified in the model (13 total). For these WDs, we reset the age following merger and use the new WD mass to recalculate the luminosity and temperature using Equations \ref{eq:radius} and \ref{eq:lum}. Of course, the true luminosities and temperatures of WD merger remnants are uncertain. However, we note that WD J0551+4135 \citep{GentileFusillo2019}, identified as a potential high-mass, $\approx1.15\,M_{\odot}$, WD candidate based on its position in the \textit{Gaia} HR diagram \citep{Gaia2018}, occupies a similar location in the HR diagram as our proposed WD merger remnants (i.e., to the upper-left of the standard WD cooling sequence). \citet{Hollands2020} argue the properties of this object are consistent with it having formed from an earlier WD merger. Although WD J0551+4135 is not itself identified in a star cluster, its observed features may provide a clue as to the properties of WD merger remnants that may potentially be observed in clusters. For reference, we show the properties of WD J0551+4135 derived from \citet{Hollands2020} as a red star in Figure \ref{fig:HR_diag}. If indeed a fraction of WD mergers result in young, massive WD remnants, we predict core-collapsed clusters like NGC 6397 may host up to $\mathcal{O}(10-20)$ such objects, which may be observationally identified by their unique position on the HR diagram.

Another possible outcome of WD mergers/collisions is NS/pulsar formation. This outcome is expected if electron capture onto Ne$^{20}$ and Mg$^{24}$ is triggered leading to collapse of the O/Ne/Mg core \citep[e.g.,][]{Miyaji1980,Nomoto1984}. In our models, we assume this process occurs when the total mass of a WD+WD dynamical collision exceeds the Chandrasekhar limit. This is almost certainly an oversimplification; the precise details of WD mergers/collisions leading to NS formation are far more complex \citep[e.g.,][]{Schwab2021}. However, our current implementation can at least roughly quantify the number of NSs/pulsars that may be formed through this process.

Pulsars are observably abundant in GCs \citep[e.g.,][]{Ransom2008}. Most pulsars in Galactic GCs have millisecond periods, unsurprising since millisecond pulsars have low spin-down rates and thus long lifetimes ($\gtrsim 10^{10}\,$yr). Meanwhile, young pulsars with periods $\gtrsim 30\,$ms have relatively short lifetimes ($\approx10^7\,$yr). Thus young pulsars formed through standard stellar evolution at early time in GCs are no longer observed at present day. However, young pulsars may form at late times through dynamical interactions, including WD mergers, NS/WD--MS star collisions, and/or partial NS recycling via accretion in binaries \citep[e.g.,][]{VerbuntFreire2014,Ye2018}. In this case, observations of young pulsars in clusters may indicate prior WD mergers in clusters.

In total, 17 young pulsars (spin period $\geq 30\,$ms) are presently observed in Galactic GCs; 8 in core-collapsed GCs and 9 non-core-collapsed GCs.\footnote{http://www.naic.edu/~pfreire/GCpsr.html}
Given that about $20\%$ of Galactic GCs are core-collapsed \citep{Trager1995}, this nearly even split tentatively suggests that young pulsars are overabundant in core-collapsed clusters. This may be due to the amplified WD merger rate in core-collapsed clusters at late times.

In \texttt{CMC}, we follow the prescriptions in \citet{Ye2018} to model pulsar magnetic fields and spin periods. When a NS forms through a WD--WD collision, we assume a new young pulsar forms. Once formed, these pulsars typically survive for $\sim 10^7-10^8\,$yr before their magnetic fields and spin periods evolve sufficiently to fall below the pulsar ``death'' line. In our best-fit models for NGC 6397, we identify $1-2$ young pulsars (defined here simply as pulsars with spin periods $\geq30\,$ms) per snapshot.
If WD binary mergers were assumed to produce pulsars in \textit{all} cases, the number of young pulsars per model would likely increase by a small factor. 

Some studies have argued that conservation of angular momentum during accretion-induced collapse may potentially lead to NSs with spin periods of a few milliseconds \citep[e.g.,][]{King2001,Dessart2006}. In this case, WD mergers may be progenitors of some fraction of the young millisecond pulsars observed in GCs. Additionally, if at least one of the merging WDs is highly magnetic \citep[e.g.,][]{Ferrario2015}, conservation of magnetic flux during accretion-induced collapse may produce NS magnetic fields of $\gtrsim 10^{14}\,$G \citep[e.g.,][]{King2001,Levan2006,Dessart2007,Schwab2021}, comparable to field strengths expected for magnetars \citep[e.g.,][]{WoodsThompson2006,OlausenKaspi2014}. Although a magnetar has yet to be observed in an old GC, WD--WD mergers may provide a possible pathway to form such objects.

Alternatively, several studies argue that the most likely outcome of a massive WD merger/collision is an explosive transient with no remnant left behind. Possible transients include standard Type Ia SNe arising from carbon detonation \citep[e.g.,][]{Webbink1984,Woosley1994,Pakmor2010,Shen2018,Sato2016,Perets2019}, subluminous transients such as SNe Iax \citep[e.g.,][]{Foley2013} that may result from failed detonation \citep[e.g.,][]{Kromer2015}, or Ca-rich gap transients \cite[e.g.,][]{Kasliwal2012} potentially formed from He shell ignition \citep[e.g.,][]{DessartHillier2015, Polin2021}. As shown in Section \ref{sec:model_rate}, we estimate an overall WD merger rate of up to roughly $50\,\rm{Gpc}^{-3}\,\rm{yr}^{-1}$ in the local universe from core-collapsed GCs. The volumetric rate of Type Ia SNe observed in the local universe is $2.5 \pm 0.5 \times 10^{4}\,\rm{Gpc}^{-3}\,\rm{yr}^{-1}$ \citep{Li2011,Cappellaro2015}. Note that SNe Iax and Ca-rich gap transient rates are comparable to the Type Ia rate, within a factor of a few \citep[e.g.,][]{Kasliwal2012,Foley2013}. So, even in the scenario where \textit{all} WD mergers in clusters lead to explosive transients, they are unlikely to constitute more than $0.1-1\%$ of the observed rate for these events in the local universe. Nonetheless, if a Type Ia SN or other transient were observed in association with a host cluster (either within the cluster or slightly offset from it if the WD binary merged after dynamical ejection), this would provide clear evidence for the dynamical processes described above.

We have suggested several possible outcomes of WD mergers/collisions and proposed some observational tests. In general, more detailed studies are required to determine more precisely the expected outcome of WD mergers/collisions and the properties of their potential remnants (if any). Some recent studies on this topic include \citet{Guerrero2004,LorenAguilar2009,Rosswog2009,Dan2011,WoosleyKasen2011,GarciaBerro2012,Shen2012,AznerSiguan2013,Piro2014,Kromer2015,Schwab2016,Polin2019,Perets2019,Schwab2021}.

\section{Collisions between white dwarfs and main sequence stars}
\label{sec:WDMS_coll}

As shown in Figure \ref{fig:radial_dist}, near the centers of core-collapsed clusters, a subdominant population of main sequence (MS) stars will mix with the massive WDs; in the case of our NGC 6397-like model, the ratio of WDs to MS stars is roughly 5:1 within the core defined at $0.07\,\rm{pc}$. Due to mass segregation, the MS population in the central regions consists preferentially of stars near the turn-off mass (roughly $0.9\,M_{\odot}$ for a cluster of this age), massive stars that formed via stellar collisions \citep[which would be potentially observed as blue stragglers; e.g.,][]{Bailyn1995}, and MS binaries. Indeed, in the NGC 6397-like model shown in Figure \ref{fig:radial_dist}, the median mass of MS stars in the central region is roughly $0.6\,M_{\odot}$ compared to $0.2\,M_{\odot}$ in the entire cluster. In addition to the role of MS stars in WD binary hardening (see Figure \ref{fig:interactions}), interactions between these two populations will cause WD--MS collisions and WD--MS binary formation, which we discuss further in Section \ref{sec:conclusion}.

In the gravitational-focusing limit, the cross section for WD--MS collisions is proportional to the MS star's radius, and is therefore roughly $100$ times larger than the WD--WD collision cross section (see Equations \ref{eq:coll_ss}-\ref{eq:coll_bs}). In this case, although MS stars are outnumbered by WDs in a core-collapsed cluster's inner region, MS--WD collisions are expected to occur more than an order of magnitude more frequently than WD--WD collisions.

In total, $4750$ WD--MS collisions occur at late times ($t>9\,\rm{Gyr}$) in our NGC 6397-like models (on average 263 per model). Of these, roughly $8\%$ and $92\%$ occur through single--single and binary-mediated resonant encounters, respectively. Performing rate calculations analogous to those in Section \ref{sec:model_rate}, we estimate a WD--MS collision rate of roughly $10^{-7}\,\rm{yr}^{-1}$ per core-collapsed cluster, roughly $f\times10^{-5}\,\rm{yr}^{-1}$ in the Milky Way (as before, $f$ is the fraction of clusters that are core-collapsed), and roughly $f\times10^3\,\rm{Gpc}^{-3}\,\rm{yr}^{-1}$ in the local universe.

In our \texttt{CMC} models, we assume a WD--MS merger produces a giant-branch star with core-mass equal to the initial WD (see Section \ref{sec:methods}). In reality, this treatment is an oversimplification and the actual outcome of WD--MS collisions is very uncertain. Using hydrodynamic models, \citet{Ruffert1992} demonstrated that, in the case of off-axis collisions ($r_{\rm{peri}}\sim r_{\rm{MS}}$), a disk may ultimately form with as little as $\sim10\%$ of the mass being expelled from the system. However, others \citep[e.g.,][]{SharaShaviv1977,SharaRegev1986,MichaelyShara2021} have argued that, in the case of head-on collisions with little angular momentum, nuclear burning ignited from a shock created as the WD passes supersonically through the MS star may produce roughly $10^{48}-10^{50}\,\rm{erg}$, comparable to the binding energy of the MS star. This suggests the majority of the disrupted MS star material becomes unbound from the system. Any unbound material is expected to expand at velocity of order the WD's escape velocity (roughly $7\times10^8\,\rm{cm\,s}^{-1}$ for a $\approx1\,M_{\odot}$ CO WD). As shown in \citet{MichaelyShara2021}, the characteristic timescale for this adiabatically expanding shell to become optically thin (roughly $10^6-10^8\,$s) can be combined with the energy generated from nuclear burning to estimate a total luminosity  $L\approx10^{48}\,\rm{erg}/(10^6-10^8\,\rm{s}) \approx 10^{40}-10^{42}\,\rm{erg\,s}^{-1}$ for these collisions, comparable in brightness to a kilonova \citep[e.g.,][]{Metzger2010} and a factor of $10-100$ times dimmer than a standard core-collapse supernova.  We reserve for future study a more detailed examination of potential transient signatures associated with these events.

Regardless of whether the disrupted stellar material ultimately remains bound to the WD following the collision, the original MS star is destroyed. Thus, the cumulative effect of these WD--MS collisions is a depletion of the MS star population. If measurable, this depletion may provide an observational test of the results presented here. In the NGC 6397-like model shown in Figure \ref{fig:radial_dist}, a total of 323 WD--MS collisions occur once the cluster has undergone core collapse. The 323 MS stars destroyed through these collisions is negligible compared to the total number of MS stars in the entire cluster (roughly $2\times10^5$ at late times). However, only 500 MS stars reside within the innermost $0.1\,$pc (within the WD subsystem; see Figure \ref{fig:radial_dist}). Of the 323 total WD--MS collisions, roughly $90\%$ occur within the innermost $0.1\,$pc. Thus, within the central region, the depletion of MS stars through WD collisions is significant.

As a result of mass segregation, we expect the most massive MS stars to preferentially reside in the center of the cluster. This includes blue straggler stars (BSSs) that were formed from previous star--star mergers \citep[e.g.,][]{Shara1999,Bailyn1995,Ferraro2012}. In total, we identify 39 BSSs\footnote{Here we define a BSS simply as any MS star with mass in excess of the MS turnoff mass ($M_{\rm{to}}\approx0.8\,M_{\odot}$ for old clusters like NGC 6397). In these models, nearly all BSSs are formed through previous stellar collisions. See \citet{Kremer2020} for further discussion of BSSs in \texttt{CMC} models.} in the best fit model for NGC 6397 (simulation \texttt{17}). In this model, we identify 94 BSS--WD collisions (roughly $30\%$ of all 323 WD--MS collisions), more than double the total number of BSSs identified in the model. We conclude that WD collisions may lead to a significant depletion of BSSs in core-collapsed clusters. To this point, \citet{Piotto2004} analyzed a catalog of nearly 3000 BSSs observed in Galactic GCs and noted that the BSS frequency in post-core-collapse GCs is comparable to the frequency in normal GCs, even though post-core-collapse cluster have relatively higher central densities. We can speculate that this may be explained in part by the regulation of BSSs through collisions with WDs. We reserve more detailed study of this possibility for future work.

Additionally, if one of the members of a MS--MS binary collides with a WD during a binary-resonant encounter, the original MS--MS binary is disrupted. In this case, we expect WD--MS collisions to reduce the number of MS--MS binaries in the cluster. In total, 82 MS--MS binaries are destroyed through WD--MS collisions in our NGC 6397-like model of interest. These 82 destroyed binaries are negligible compared to the total number of MS--MS binaries in the entire cluster (roughly $10^4$ for our models with $5\%$ primordial binary fraction). However, only 76 MS--MS binaries reside within the innermost $0.1\,$pc, indicating that in the cluster's center, WD--MS collisions destroy a significant fraction of MS--MS binaries.

\begin{figure*}
    \includegraphics[width=\linewidth]{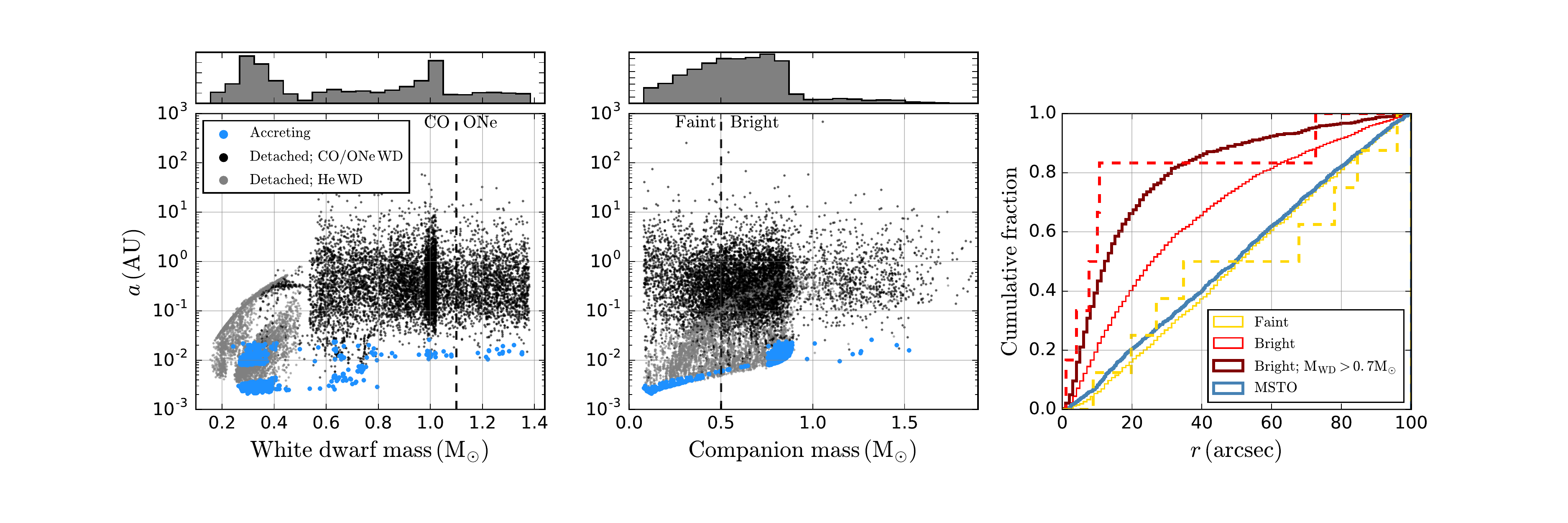}
    \caption{\footnotesize All WD--MS binaries identified at late times in our NGC 6397-like models. \textit{Left and middle panels:} orbital separation versus WD mass (left) and MS companion mass (middle) . In these panels, black (gray) circles indicate systems with a CO/ONe (He) WD. Blue circles indicate accreting systems. We denote as vertical dashed lines the boundary in WD mass separating CO and ONe WDs (left panel) and the boundary in MS mass separating our definitions of ``faint" versus ``bright" systems, following \citet{Cohn2010}. \textit{Right panel:} cumulative fraction of accreting WD--MS binaries within the inner $100\,$arcsec in our models (solid lines) compared to the distributions inferred from observed CVs in \citet{Cohn2010} (dashed lines). For reference, we also include the distribution of main-sequence turn-off (MSTO) stars (blue curve).}
    \label{fig:WDMS}
\end{figure*}

\section{Discussion and conclusions}
\label{sec:conclusion}

\subsection{Summary}

\begin{enumerate}
    \item Using a set of $N$-body models that effectively match the present observed properties of NGC 6397, we demonstrate this cluster likely contains a central population of $\sim10^3$ WDs ($\sim10^3\,M_{\odot}$ in total mass) within the cluster's innermost $0.07\,\rm{pc}$. This supports the earlier finding of \citet{Rui2021b} and provides a natural explanation for the dark diffuse population reported in NGC 6397 by \citet{Vitral2021}.
    \item We argue that such WD-dominated cores with central densities $\geq10^6\,\rm{pc}^{-3}$ are a feature of all core-collapsed clusters at present day. Once all stellar-mass BHs have been dynamically ejected, the cluster core is no longer supported against collapse and lower-mass stellar populations (in particular, CO and ONe WDs) efficiently segregate to the cluster center. Ultimately, ``burning" of central WD binaries halts the core collapse process.
    \item The high central densities of WDs in core-collapsed clusters facilitate formation and dynamical hardening of WD binaries. Ultimately these binaries will merge through either GW inspiral or direct collision with other WDs. We predict a WD merger rate of up to roughly $f\times50\,\rm{Gpc}^{-3}\,\rm{yr}^{-1}$ in the local universe, where $f$ is the fraction of all globular clusters that have undergone core collapse. For Milky Way globular clusters, $f \approx 0.2$.
    \item Prior to merger, inspiraling WD binaries will be observable as GW sources. Over its anticipated $5\,\rm{yr}$ mission lifetime, we estimate LISA may resolve of order one WD--WD binary at millihertz frequencies in NGC 6397 and of order a few WD--WD binaries in Galactic core-collapsed clusters overall. Proposed decihertz GW detectors like DECIGO may resolve GW merger signals from dozens of these systems in clusters out to distances of roughly $1\,\rm{Gpc}$.
    \item The vast majority ($\gtrsim90\%$) of all WD mergers identified in our models have total mass greater than the Chandrasekhar limit. The outcome of these WD mergers is uncertain, with some possibilities including an explosive transient like a Type Ia SN, formation of a young massive WD, or NS formation via accretion-induced collapse. At most, WD mergers/collisions from clusters may constitute roughly $0.1-1\%$ of the overall Type Ia SN rate. If a fraction of mergers lead to massive WD remnants, these objects may be identifiable by their unique location on the HR diagram to the left of the standard WD cooling sequence. Alternatively, if WD mergers/collisions produce NSs, these NSs may be detectable as young pulsars/millisecond pulsars or magnetars in old GCs.
    \item Finally, we show dynamical interactions between WDs and MS stars in core-collapsed clusters lead to WD--MS collisions occurring at a rate of up to $f\times1000\,\rm{Gpc}^{-3}\,\rm{yr}^{-1}$ in the local universe. Although the exact outcome of these collisions is uncertain, previous studies have shown they may be observed as bright transients with peak luminosities comparable to kilonovae. Regardless of the exact outcome, we predict WD--MS collisions reduce the number of blue straggler stars and MS--MS binaries by a factor of a few within the innermost regions of core-collapsed clusters.
\end{enumerate}

\subsection{WD--main sequence star binaries}
\label{sec:WDMS_bin}

Also relevant is the formation of WD--MS binaries, some of which may be hardened to the point of Roche lobe overflow. In the left two panels of Figure \ref{fig:WDMS}, we show all distinct WD--MS binaries found at late times in our NGC 6397-like models. Blue points denote accreting systems while black and gray points indicate detached systems with CO/ONe WDs and He WDs, respectively. On average, we identify $287$ detached WD--MS binaries and $23$ accreting systems per model at any given snapshot in time in our NGC 6397-like models. We find that $65\%$ of the detached systems, but only $25\%$ of the accreting systems, form through dynamical interactions rather than as primordial binaries.

Observationally, these accreting binaries are expected to be identified as cataclysmic variables (CVs). CVs are well observed in GCs \citep[e.g.,][]{Knigge2012} and also well-studied theoretically \citep[e.g.,][]{Grindlay1995,Ivanova2006}. In particular, \citet{Belloni2016,Belloni2017a,Belloni2017b,Belloni2019} have studied CVs in details using the MOCCA-Survey Database of GC models. The studies by Belloni et al. have demonstrated, among many other points, that the detectable CV population in clusters is composed primarily of CVs formed from primordial binaries (with most being formed through common envelope evolution) and in fact dynamical interactions are more likely to break apart primordial CV progenitors than lead to dynamical formation of new CVs. This is consistent with the findings of \citet{Kremer2020} and with this study, where (as noted above) only a small fraction of accreting WD--MS binaries in our NGC 6397-like models form dynamically. Additionally, these studies have demonstrated that only a few percent of all CVs in a particular GC are likely to be detectable and that high primordial binary fractions \citep[see][]{Kroupa2013,Belloni2017c} are necessary to reproduce the observed number of CVs in NGC 6397 and other clusters.

In NGC 6397 specifically, 15 CVs have been observed \citep{Cohn2010}. In \citet{Rui2021}, we pointed out that our best-fit model for NGC 6397 contains a comparable number of accreting WD--MS binaries. In \citet{Cohn2010}, the observed sample of CVs is divided into ``faint'' (9 total) and ``bright'' (6 total) categories, indicative of the apparent magnitude of each object. As pointed out in \citet{Cohn2010}, the brightness of each system is determined by the mass of the MS companion. In our models, we use a MS mass of $0.5\,M_{\odot}$ to separate the faint and bright systems. At a distance of $2.3\,$kpc, a $0.5\,M_{\odot}$ MS star would be observed with an apparent V-band magnitude of roughly $+21\,$mag, consistent with the faint/bright boundary in \citet{Cohn2010}; see their Figure 3. Of the roughly $23$ accreting WD--MS binaries found on average in our models, roughly $16$ and $7$ fall into the faint and bright categories, roughly consistent with the observed numbers in \citet{Cohn2010}. 

In the right-hand panel of Figure \ref{fig:WDMS}, we show the radial distribution of our faint and bright model CVs within the inner $100\,$arcsec of our models compared to the radial distributions of observed systems published in \citet{Cohn2010}. Solid lines denote model distributions while dashed lines denote observed distributions. For reference, we also show as a blue curve the radial distribution for
main-sequence turn off (MSTO) stars (defined simply as any MS star with mass in the range $[0.9,1] \times M_{\rm{to}}$, where $M_{\rm{to}}$ is the turn-off mass). 

The radial distribution of faint CVs matches well both the observed distribution and the MSTO distribution, consistent with the results of \citet{Cohn2010}. Furthermore, the bright CVs (pale solid red curve) in our models are more centrally concentrated than the faint CVs. Previous studies \citep[e.g.,][]{Nelemans2016} have argued that nova eruptions prevent stable accretion and CV formation for low-mass WDs ($M\lesssim0.7\,M_{\odot}$), motivated in part by the fact that WDs observed in CVs are generally high-mass \citep[e.g.,][]{Zorotovic2011}. Along these lines, we show as a dark red curve the distribution of bright CVs excluding the systems with WD masses $<0.7\,M_{\odot}$. As shown, in this case, the bright population matches much more closely the observed distribution of bright CVs (dashed red curve). Finally, note that of the roughly $7$ bright accreting WD--MS binaries found per model, roughly $2$ of these on average feature a WD of mass $0.7\,M_{\odot}$.

The agreement in our models with observations of overall numbers and radial distributions of CV candidates is satisfying. However, we emphasize the physical processes relevant to CVs and their potential detectability are far more complex than considered here. We reserve more detailed study of accreting WD--MS binaries for future work.

\subsection{Effect of primordial binary fraction}
\label{sec:primordial_binaries}

\begin{figure}
    \includegraphics[width=\linewidth]{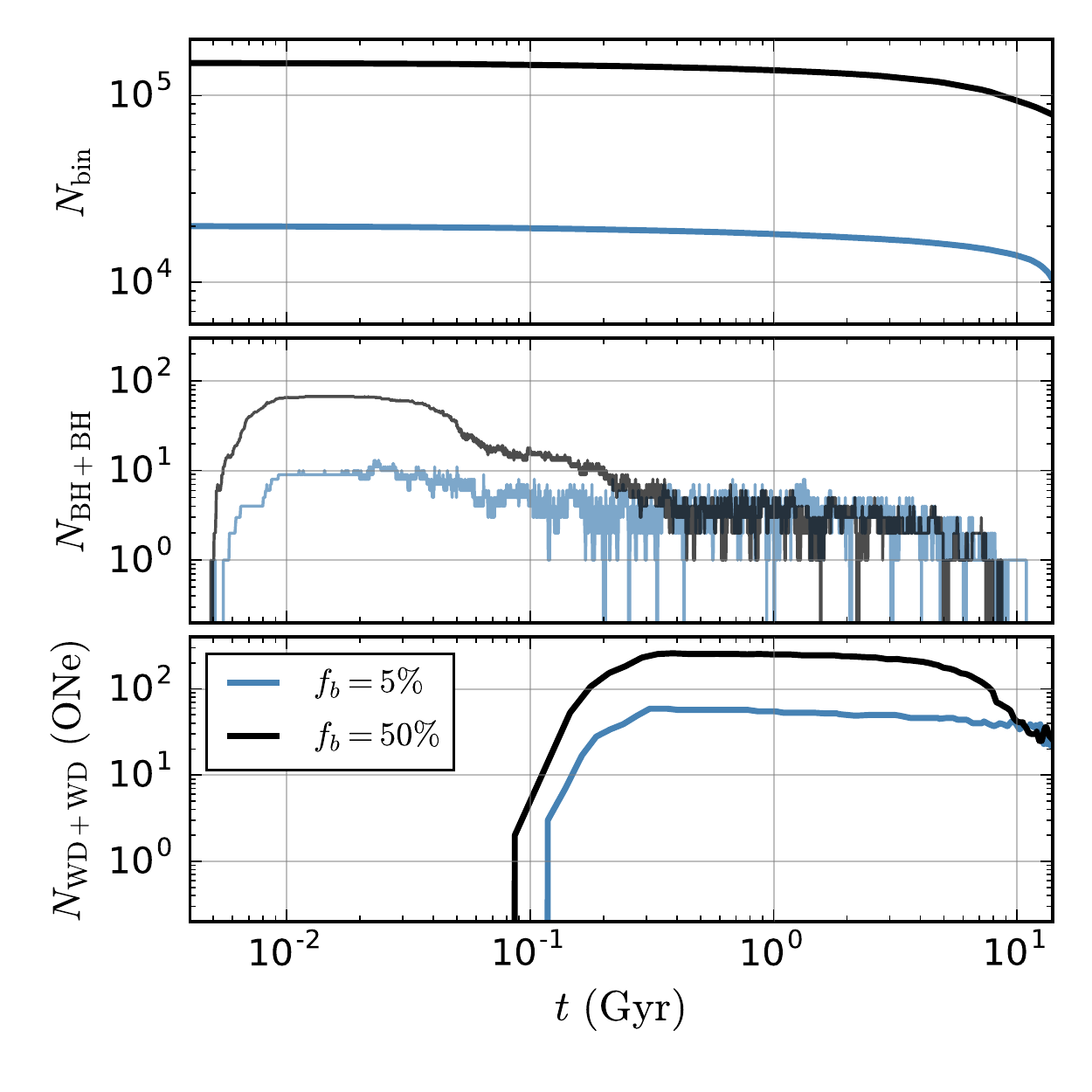}
    \caption{\footnotesize Total number of binaries of all types (top panel), total number of BH--BH binaries (middle), and total number of massive WD binaries (here defined simply as ONe WDs; bottom panel) versus time for model \texttt{21} (primordial binary fraction of $50\%$) and model \texttt{10} ($f_b=5\%$). Populations that are dominated by dynamics (BHs at early times when a BH subsystem is present and WDs at late times when a WD subsystem forms) converge to the same binary fraction, independent of the assumed primordial binary fraction.}
    \label{fig:bin}
\end{figure}

Here we explore how the results of this study may depend upon the primordial binary fraction in clusters. Stellar-mass BH binaries are expected to form primarily through dynamical processes regardless of the details assumed about the primordial binary population \citep[e.g.,][]{Chatterjee2017a}. However, it is not immediately clear whether the primordial binary fraction may play a more prominent role in the WD case. We include an additional model (\texttt{21} in Table \ref{table:models}) assuming a $50\%$ primordial binary fraction. We summarize here the primary differences between this model and our fiducial primordial binary fraction of $5\%$.

Although the \textit{total} number of WDs remains largely unchanged, the number of He WDs does change markedly with the binary fraction. This is unsurprising given that He WDs form exclusively through binary evolution processes (see Section \ref{sec:methods}). When the primordial binary fraction increases by a factor of 10 from $5\%$ to $50\%$, the number of He WDs also increases by roughly a factor of 10 (see Table \ref{table:models}).

We find the number of WD mergers does not change significantly with binary fraction (columns 10-12 of Table \ref{table:models}). Additionally, the total number of WD--MS collisions is largely unchanged; we identify $316$ collisions in model \texttt{21} compared to an average of $263$ for the models with $5\%$ primordial binary fraction. This indicates the rates of WD mergers and WD--MS collisions are dependent primarily on the high densities of WDs achieved in the centers of core-collapsed clusters.

We explore this point further in Figure \ref{fig:bin}. Here we plot versus time the total number of binaries of all types (top panel), the total number of BH--BH binaries (middle panel), and the total number of massive WD--WD binaries (bottom panel). We show the evolution for model \texttt{21} with $f_b=50\%$ (black curves) and model \texttt{10} with $f_b=5\%$ (blue curves). Note that all other cluster parameters, including the total cluster mass, are identical in these two models. We focus on massive WDs (in this plot, defined simply as the ONe WDs) to explore specifically the \textit{dynamically active} WDs that dominate WD subsystems (see Figure \ref{fig:radial_dist}). As shown, the overall number of binaries (the majority of which are low-mass binaries that live dynamically-unperturbed in the cluster halo) remains markedly different throughout the full cluster evolution for these two primordial binary fraction assumptions. However, for populations shaped significantly by dynamics, the total number of binaries (and thus binary fraction) converges to a single value, independent of the assumed primordial binary fraction. Notably, we show this is relevant for BHs at early times ($t\lesssim10\,$Gyr), while a BH subsystem is present, and for massive WDs at late times ($t\gtrsim10\,$Gyr), once a dense WD subsystem has formed. We reserve a more detailed exploration of the dynamical processes that may explain these features for future work.

We do find that the numbers of accreting and detached WD--MS binaries are sensitive to the primordial binary fraction, since the majority of these systems are dynamically-inactive low-mass binaries. In our $f_b=50\%$ model, we find on average $1670$ detached WD--MS binaries at late times and on average $162$ accreting systems (117 and 45 of which fall into the faint and bright categories described in Section \ref{sec:WDMS_bin}). Given that a large fraction of these binaries found in the models at late times are primordial, it is not surprising that a factor of 10 increase in primordial binary fraction yields a roughly factor of 10 increase in the late time binary counts. As discussed in \citet{Belloni2019}, only a few percent of accreting WD--MS binaries in a given cluster may in fact be observed as CVs when taking into account observational considerations. In this case, a relatively large primordial binary fraction may in fact be necessary to produce CV counts on par with those observed in NGC 6397 (and in other clusters), as is pointed out by Belloni et al. Again, we reserve for future work a more detailed examination of the observable features of the CV candidates in our \texttt{CMC} models. 

We show more detailed information on the total numbers of WD--WD and WD--MS binaries of various types found in each simulation in Tables \ref{table:bin_WDWD} and \ref{table:bin_WDMS} in the Appendix.

\subsection{Future work}

\citet{Weatherford2018,Weatherford2020} showed that mass segregation measurements constrain the total number and total mass of BH subsystems in globular clusters. This result arises directly from the connection between BH populations and the overall cluster dynamics. As we have shown here (specifically see Figure \ref{fig:rho_vs_BH}), WD subsystems are also tracers of cluster dynamics, so mass segregation measurements may similarly be used to predict WD subsystems in GCs. Specifically, we expect clusters exhibiting more mass segregation \citep[parameterized with the metric $\Delta$ in][]{Weatherford2020} feature relatively dense central WD populations. In \citet{Weatherford2020}, we predicted details of BH subsystems in roughly 50 GCs observed in the ACS Survey. A future study may place similar constraints upon WD subsystems in these specific clusters or others. While we have focused here on rates of WD mergers and other transients in NGC 6397 as a representative example of a core-collapsed cluster, detailed predictions of WD populations in core-collapsed clusters more broadly would enable more precise estimates of the rates predicted here. Of course, a larger set of cluster models spanning a broader range in cluster masses and other properties would be ideal for such a study.

We considered here direct collisions between WDs (i.e., $r_p<R_1+R_2$). However, more distant close encounters in the tidal disruption or tidal capture regime may also lead ultimately to a WD merger. The minimum pericenter distance for a pair of WDs to be tidally captured may be a few$-10$ times larger than the physical WD radius \citep[e.g.,][]{Samsing2017}. Thus, if tidal capture were included (for both single--single and binary-mediated encounters), the WD merger rate estimated in Section \ref{sec:model_rate} may increase to roughly $\mathcal{O}(100)\,\rm{Gpc}^{-1}\,\rm{yr}^{-1}$. The inclusion of tidal capture may similarly increase the rates of WD--MS collisions described in Section \ref{sec:WDMS_coll}. Of course, the tidal capture process is highly uncertain and depends on physics beyond the scope of the current version of \texttt{CMC}. We reserve study of this possibility for future work.

\acknowledgements{
K.K. thanks Ilaria Caiazzo, Kevin Burdge, Jim Fuller, and Tony Piro for helpful discussions. We also thank the anonymous referee for helpful suggestions and careful review of the manuscript. K.K. is supported by an NSF Astronomy and Astrophysics Postdoctoral Fellowship under award AST-2001751.
N.Z.R. acknowledges support from the Dominic Orr Graduate Fellowship. N.C.W acknowledges support from the CIERA Riedel Family Graduate Fellowship.
S.C. acknowledges support of the Department of Atomic Energy, Government of India, under project no. 12-R\&D-TFR-5.02-0200.
F.A.R. and C.S.Y. acknowledge support from NSF Grant AST-1716762 at Northwestern University.
}

\bibliographystyle{aasjournal}
\bibliography{mybib}

\appendix

In Table \ref{table:bin_WDWD}, we show the average number of WD--WD binaries of various types found in all late-time cluster ($t>10\,$Gyr) snapshots for all simulations. In parentheses, we show the average number of each binary type found within the innermost $0.1\,$pc (see Figure \ref{fig:radial_dist}). Table \ref{table:bin_WDMS} shows the average number of WD--MS binaries of various types for each simulation, again limited only to late-time snapshots. Columns 2-4 show average binary counts for detached and accreting systems combined, while column 5-7 shows specifically the counts for accreting systems (as discussed in Section \ref{sec:WDMS_bin}).

\newpage

\begin{deluxetable}{l|c|cccccc}
\tabletypesize{\footnotesize}
\tablewidth{0pt}
\tablecaption{WD--WD Binaries \label{table:bin_WDWD}}
\tablehead{
	\colhead{$^1$} &
	\colhead{$^2$MS+MS} &
	\colhead{$^3$WD+WD} &
	\colhead{$^4$He+CO} &
	\colhead{$^5$He+ONe} &
	\colhead{$^6$CO+CO} &
	\colhead{$^7$CO+ONe} &
	\colhead{$^8$ONe+ONe}
}
\startdata
\texttt{1} & 11429.41 (66.18) & 166.91 (31.32) & 51.77 (2.09) & 1.95 (0.86) & 58.45 (12.82) & 22.32 (11.73) & 5.55 (3.59) \\
\texttt{2} & 12574.53 (57.29) & 190.41 (28.29) & 66.06 (3.53) & 0.18 (0.0) & 77.53 (11.82) & 20.24 (8.65) & 5.65 (3.65) \\
\texttt{3} & 12053.24 (60.35) & 154.12 (31.76) & 58.47 (3.76) & 2.24 (0.71) & 60.76 (15.71) & 14.29 (8.0) & 6.47 (3.47) \\
\texttt{4} & 12086.5 (66.2) & 143.3 (27.4) & 54.9 (5.0) & 1.25 (0.6) & 52.3 (12.2) & 16.55 (7.8) & 3.55 (1.8) \\
\texttt{5} & 10731.55 (55.38) & 174.12 (23.89) & 58.28 (2.71) & 0.0 (0.0) & 65.68 (11.4) & 21.29 (7.32) & 4.83 (2.09) \\
\texttt{6} & 12945.31 (38.92) & 225.0 (23.38) & 71.08 (2.08) & 0.92 (0.08) & 100.23 (11.31) & 27.38 (7.77) & 3.92 (1.85) \\
\texttt{7} & 11808.33 (73.1) & 158.33 (28.43) & 63.38 (5.43) & 2.0 (0.43) & 44.0 (9.29) & 15.43 (8.43) & 6.48 (4.14) \\
\texttt{8} & 12013.2 (69.3) & 125.8 (24.65) & 60.1 (4.6) & 1.85 (0.75) & 34.95 (9.9) & 12.6 (6.0) & 4.45 (2.8) \\
\texttt{9} & 11378.28 (60.38) & 192.26 (27.51) & 69.96 (3.64) & 1.0 (0.34) & 78.47 (13.83) & 17.09 (6.4) & 5.57 (3.02) \\
\texttt{10} & 11917.93 (62.43) & 149.75 (25.75) & 65.04 (4.71) & 1.93 (0.21) & 46.43 (11.29) & 13.32 (6.64) & 4.36 (2.68) \\
\texttt{11} & 12704.06 (37.18) & 217.41 (20.29) & 73.0 (2.06) & 0.94 (0.35) & 95.82 (9.0) & 26.12 (6.29) & 5.12 (2.35) \\
\texttt{12} & 12455.19 (44.08) & 228.7 (21.16) & 70.51 (2.43) & 2.76 (0.27) & 97.57 (10.27) & 23.95 (6.32) & 4.62 (1.7) \\
\texttt{13} & 12484.5 (51.67) & 201.78 (25.94) & 76.89 (4.0) & 1.72 (0.11) & 68.94 (9.94) & 26.5 (9.11) & 6.89 (2.56) \\
\texttt{14} & 13415.45 (59.7) & 201.35 (28.6) & 67.55 (3.75) & 1.35 (0.2) & 83.95 (14.1) & 24.25 (7.45) & 5.25 (2.9) \\
\texttt{15} & 12253.53 (68.47) & 169.0 (32.76) & 67.29 (4.18) & 0.53 (0.12) & 54.0 (12.94) & 23.65 (11.35) & 6.29 (3.88) \\
\texttt{16} & 12832.91 (61.45) & 175.36 (29.5) & 72.32 (4.73) & 2.36 (0.5) & 53.5 (12.5) & 16.05 (8.77) & 5.77 (2.41) \\
\texttt{17} & 12041.12 (61.59) & 177.76 (28.29) & 64.12 (3.71) & 1.47 (0.76) & 57.76 (11.53) & 21.76 (9.65) & 4.88 (2.35) \\
\texttt{18} & 13267.29 (37.48) & 225.38 (16.81) & 79.76 (2.38) & 2.24 (0.14) & 98.43 (7.57) & 23.52 (5.29) & 6.48 (1.33) \\
\hline
\texttt{19} & 14590.73 (0.45) & 45.45 (0.0) & 5.73 (0.0) & 0.0 (0.0) & 9.36 (0.0) & 2.36 (0.0) & 0.0 (0.0) \\
\texttt{20} & 14783.0 (0.07) & 130.87 (0.0) & 15.87 (0.0) & 0.0 (0.0) & 41.27 (0.0) & 1.0 (0.0) & 0.0 (0.0) \\
\hline
\texttt{21} & 82915.06 (131.18) & 1018.0 (33.12) & 395.76 (4.76) & 7.71 (0.82) & 422.24 (21.53) & 18.24 (3.65) & 3.12 (1.18) \\
\enddata
\tablecomments{Average number of WD--WD binaries of various types found in late time ($t>10\,$Gyr) snapshots for all models. The numbers in parentheses show average number of binaries within $0.1\,$pc (i.e., within the WD subsystem).}
\end{deluxetable}

\newpage

\begin{deluxetable}{l|cccccc}
\tabletypesize{\footnotesize}
\tablewidth{0pt}
\tablecaption{WD--MS Binaries \label{table:bin_WDMS}}
\tablehead{
    \multicolumn{1}{c}{} &
    \multicolumn{3}{c}{WD+MS binaries} &
    \multicolumn{3}{c}{Accreting WD+MS} \\
    \colhead{$^1$} &
	\colhead{$^2$He+MS} &
	\colhead{$^{3}$CO+MS} &
	\colhead{$^{4}$ONe+MS} &
	\colhead{$^{5}$All}&
	\colhead{$^{6}$Faint}&
	\colhead{$^{7}$Bright}
}
\startdata
\texttt{1} & 170.77 (3.36) & 76.55 (12.95) & 14.05 (6.59) & 18.82 (1.64) & 12.36 (0.0) & 6.45 (1.64) \\
\texttt{2} & 209.71 (4.82) & 91.94 (15.47) & 12.35 (3.59) & 19.59 (1.59) & 11.71 (0.06) & 7.88 (1.53) \\
\texttt{3} & 199.59 (4.59) & 75.06 (12.41) & 13.41 (5.24) & 18.41 (0.65) & 14.82 (0.12) & 3.59 (0.53) \\
\texttt{4} & 191.75 (4.95) & 91.2 (15.6) & 12.3 (4.4) & 24.55 (1.5) & 19.1 (0.15) & 5.45 (1.35) \\
\texttt{5} & 192.15 (3.66) & 95.75 (11.57) & 16.66 (4.11) & 27.23 (1.28) & 17.05 (0.11) & 10.18 (1.17) \\
\texttt{6} & 240.92 (1.77) & 111.38 (10.08) & 12.69 (3.54) & 30.92 (0.23) & 22.85 (0.15) & 8.08 (0.08) \\
\texttt{7} & 145.43 (3.43) & 78.9 (15.1) & 13.67 (6.0) & 16.38 (0.86) & 14.1 (0.38) & 2.29 (0.48) \\
\texttt{8} & 133.75 (3.55) & 71.65 (11.9) & 17.4 (7.0) & 17.85 (1.55) & 11.0 (0.1) & 6.85 (1.45) \\
\texttt{9} & 154.81 (2.85) & 118.68 (13.55) & 15.53 (5.34) & 19.68 (0.62) & 15.77 (0.13) & 3.91 (0.49) \\
\texttt{10} & 132.89 (3.11) & 72.93 (12.46) & 15.82 (6.36) & 15.96 (1.07) & 10.61 (0.07) & 5.36 (1.0) \\
\texttt{11} & 198.82 (2.24) & 100.0 (9.12) & 14.76 (3.82) & 16.53 (0.06) & 12.59 (0.0) & 3.94 (0.06) \\
\texttt{12} & 181.3 (2.92) & 111.68 (9.08) & 20.46 (3.7) & 19.7 (0.84) & 11.81 (0.08) & 7.89 (0.76) \\
\texttt{13} & 225.33 (3.61) & 98.11 (12.5) & 18.83 (5.5) & 33.06 (1.33) & 22.72 (0.44) & 10.33 (0.89) \\
\texttt{14} & 260.45 (4.95) & 110.9 (11.8) & 17.15 (5.1) & 31.2 (1.2) & 21.5 (0.1) & 9.7 (1.1) \\
\texttt{15} & 232.06 (3.88) & 89.88 (13.53) & 14.76 (5.65) & 31.18 (0.82) & 21.35 (0.24) & 9.82 (0.59) \\
\texttt{16} & 233.32 (5.05) & 102.45 (15.64) & 12.59 (4.0) & 32.0 (2.23) & 22.14 (0.45) & 9.86 (1.77) \\
\texttt{17} & 237.0 (5.59) & 102.94 (14.76) & 14.47 (4.94) & 33.82 (1.06) & 25.53 (0.06) & 8.29 (1.0) \\
\texttt{18} & 255.71 (2.38) & 110.9 (6.9) & 19.57 (3.52) & 37.52 (0.62) & 27.38 (0.05) & 10.14 (0.57) \\
\hline
\texttt{19} & 303.0 (0.0) & 92.73 (0.0) & 3.82 (0.0) & 52.09 (0.0) & 27.18 (0.0) & 24.91 (0.0) \\
\texttt{20} & 647.93 (0.07) & 164.33 (0.0) & 4.0 (0.0) & 68.07 (0.0) & 47.87 (0.0) & 20.2 (0.0) \\
\hline
\texttt{21} & 1450.35 (8.29) & 413.29 (16.94) & 36.24 (6.82) & 174.88 (2.47) & 128.76 (0.53) & 46.12 (1.94) \\
\enddata
\tablecomments{Average number of WD--MS binaries of various types found in late time ($t>10\,$Gyr) snapshots for all models. The numbers in parentheses show average number of binaries within $0.1\,$pc.}
\end{deluxetable}

\end{document}